# The Good, the Bad, and the Monstrous: Predicting Highly Change-Prone Source Code Methods at Their Inception

SHAIFUL CHOWDHURY, SQM Research Lab, Computer Science, University of Manitoba, Canada

The cost of software maintenance often surpasses the initial development expenses, making it a significant concern for the software industry. A key strategy for alleviating future maintenance burdens is the early prediction and identification of change-prone code components, which allows for timely optimizations. While prior research has largely concentrated on predicting change-prone files and classes—an approach less favored by practitioners—this paper shifts focus to predicting highly change-prone methods, aligning with the preferences of both practitioners and researchers. We analyzed 774,051 source code methods from 49 prominent open-source Java projects. Our findings reveal that approximately 80% of changes are concentrated in just 20% of the methods, demonstrating the Pareto 80/20 principle. Moreover, this subset of methods is responsible for the majority of the identified bugs in these projects. After establishing their critical role in mitigating software maintenance costs, our study shows that machine learning models can effectively identify these highly change-prone methods from their inception. Additionally, we conducted a thorough manual analysis to uncover common patterns (or concepts) among the more difficult-to-predict methods. These insights can help future research develop new features and enhance prediction accuracy.

## 1 INTRODUCTION

The cost of software maintenance has long been a concern for the software industry [46]. Maintenance cost may often exceed the original development cost [14]—so much so that it can account for up to 90% of the total ownership cost [100]. As such, a significant amount of research has been dedicated to understanding and predicting maintenance-prone code components [16, 17, 24, 25, 50, 51, 68, 81]. In general, the change-proneness [7, 23, 25, 98] and bug-proneness [24, 68, 81] of a code component are considered two of the most significant indicators of future software maintenance cost and effort: a code component that changes a lot or is frequently associated with bug-fix commits is generally more maintenance-prone. Therefore, building models to predict and improve change- and bug-prone components in their early lifetime would significantly help reduce future maintenance burdens. Developing bug-prediction models, however, is considered extremely challenging due to the lack of accurate and noise-free bug datasets [24, 37, 38]. Researchers often label all modified code in a bug-fix commit as bug-prone, a practice that can be flawed due to the presence of tangled changes—unrelated code modifications made and committed simultaneously [24, 37, 52]. Reducing noise from bug datasets by mitigating the impact of tangled code changes remains an ongoing and active research area [24, 36].

In this paper, we, therefore, focus on characterizing and predicting change-prone code components. While one might argue that change itself is not inherently harmful—since software must evolve to accommodate updates in requirements and features—software systems typically grow through the addition of new methods and classes rather than frequent modifications to existing code [108]. Also, frequent changes to an existing code component violate the open-closed principle [74], which advocates that software entities should be open for extension but closed for modification. This principle, when violated, can be detrimental to software maintenance. While change-proneness remains a valuable predictor of maintenance future burdens, it can also serve as an indicator of potential bug-prone components: code that experiences frequent changes is often more susceptible to bugs [7, 85].

Author's address: Shaiful Chowdhury, shaiful.chowdhury@umanitoba.ca, SQM Research Lab, Computer Science, University of Manitoba, Winnipeg, MB, Canada.

The significance of developing models to predict change-prone code components has been widely acknowledged and extensively studied by the research community [7, 16, 17, 50, 51, 53, 62, 90]. However, these models typically operate at the class level, which can be challenging for practitioners who find it difficult, laborious, and time-consuming to manage maintenance tasks at such a granularity [35, 68, 81]. Motivated by this observation, our paper shifts the focus to characterizing and predicting change-proneness at the method level. We collected change histories for an extensive dataset of 774,051 Java methods from 49 widely-used open-source projects, using the advanced CodeShovel tool [34, 35]. With this large dataset, we aim to determine whether changes occur predominantly within a small fraction of methods and whether these methods are responsible for the majority of bugs in these projects. Additionally, we evaluate the effectiveness of various machine learning models in identifying these highly maintenance-prone methods. Specifically, we answer four research questions that represent the core contributions of our paper.

**RQ1:** Does change-proneness at the method-level granularity follow the Pareto 80/20 principle?

**Contribution 1:** We demonstrate that, in most projects, approximately 80% (or more) of the changes originate from about 20% of the methods. We refer to these 20% of methods as *monstrous* because they undergo frequent modifications that significantly impact software maintenance. Conversely, we observed that on average, around 43% of methods per project remain unchanged. These methods are termed *good* as they exemplify best practices for writing code to minimize future maintenance costs and effort. Methods that are neither *good* nor *monstrous* are classified as *bad* methods.

**RQ2:** What percent of bugs can be detected by capturing the *monstrous* methods?

**Contribution 2:** We found that by capturing the top 20% change-prone methods (i.e., *monstrous* methods), we can capture up to 80% of the bugs in a project. This is encouraging because it suggests that change-prediction models can also effectively identify bug-prone methods.

**RQ3:** Can we use machine learning models to accurately predict the *monstrous* methods when they are born?

**Contribution 3:** We first assessed whether a single popular code metric could effectively identify the *monstrous* methods. Our results indicate that metrics such as size or complexity (e.g., McCabe) alone are insufficient for capturing monstrous methods, as the rank correlation between size (or complexity) and change-proneness is only weak to moderate. Subsequently, we demonstrated that by using 17 relevant predictors (e.g., size, complexity, readability, fanout, etc.), various machine learning models can be developed to predict the *monstrous* methods from their inception. Early detection of maintenance-prone methods as they are introduced into a system can be instrumental in reducing future maintenance efforts.

**RQ4:** When do the machine learning models fail to detect the *monstrous* methods?

**Contribution 4:** Although the prediction accuracy[1] of our models is encouraging (e.g., precision 75% and recall 80%), we sought to understand when a machine learning model fails to predict a highly change-prone method. We performed a qualitative analysis on 100 selected method samples from our dataset: 50 samples, termed *surprisingly good methods*, were methods that never changed despite having high code complexity, size, etc.; the remaining 50 samples, termed *surprisingly monstrous* methods, were methods that experienced frequent changes despite being small and less complex. We identified several common patterns and characteristics that help explain these unexpected results. For instance, a small method might undergo numerous changes if it was an incomplete implementation or contained technical debt. These insights could be valuable for developing new code metrics and enhancing future change prediction models.

To enable replication and extension, we share our dataset publicly.[2]

[1] in this paper, unless otherwise stated, accuracy indicates precision, recall, and F-measure

[2] https://github.com/shaifulcse/good_bad_monstrous_public_data/

**Paper organization.** In Section 2, we discuss the related work and the motivations of this study. Section 3 discusses the methodology. Section 4 shows the findings of our experiments. In Section 5, we discuss and summarize our findings, implications for researchers and practitioners, and the threats that can impact our results. Section 6 concludes this paper.

## 2 RELATED WORK & MOTIVATION

In this section, we discuss the related studies on software maintenance. We also discuss how these previous studies motivated and guided the objectives of our study.

### 2.1 Maintenance, Bug-proneness, and Change-proneness

Software maintenance and quality have been extensively studied for over forty years [56], with change-proneness and bug-proneness historically recognized as crucial maintenance indicators. Researchers have developed models to predict components that are more susceptible to these indicators to optimize resources and enable early interventions through code review [118], testing [19], and code refactoring [103]. Change prediction and bug prediction research have generally focused on class/file level granularity. Koru *et al.* [53] applied tree-based models and observed that code metrics are useful for understanding change-prone classes. Khomh *et al.* [50] experimented with 29 code smells and found that classes with code smells are generally more change-prone than classes without any code smells. Romano *et al.* [90] found code metrics to be useful in predicting change-prone Java interfaces. Similar to change-proneness prediction models, bug prediction models [2, 4, 10, 12, 31, 49, 72, 121, 122], aim to identify classes or files likely to be associated with future bugs.

However, class-level bug and change prediction models have limited practical utility [35, 81, 98]. For example, most methods within a class are generally bug-free [68]. As a result, practitioners may need to invest significantly more time in locating bugs when relying on class-level predictions.

To alleviate the issues of class-level bug prediction models, method-level bug prediction has been studied significantly in recent research [29, 30, 68, 81, 102]. Except for the study of Pascarella *et al.* [81], all other studies claimed excellent performance of method-level bug prediction models. However, a recent study by Chowdhury *et al.* [24] found that high prediction accuracy in the previous method-level bug prediction research was inflated due to unrealistic evaluation scenarios. When these models were tested with realistic scenarios—e.g., no future data was used while training the models—the accuracy of all those models became extremely poor. The authors found that the biggest obstacle in bug prediction research is the lack of accurately labeled data sets, mainly due to tangled changes that developers often commit. Method-level bug prediction, therefore, remains an open and challenging research problem [24, 81].

**Motivation.** In this paper, we focus on developing method-level change prediction models, which, despite their importance, are understudied. Predictions at a finer granularity, such as line level, are prone to inaccuracies because similar lines may appear by chance, leading to numerous false positives [35, 96, 110]. With accurate method-level change history tracing tools like CodeShovel [35], we can create a precise method-level change history dataset. Although change-proneness is a crucial maintenance indicator, we hypothesize—and later demonstrate—that identifying highly change-prone methods can also help in detecting highly bug-prone methods.

### 2.2 Maintenance Predictors

Previous studies exploited different product metrics [17, 29, 50, 51, 68, 81, 90, 122], process metrics [32, 68, 71, 81], and developer-centric metrics [16, 26] to build maintenance prediction models. Product metrics are generally related to the characteristics of source and test code. At the system level, some popular product metrics are decoupling level [67], propagation cost [60], and independence level [97]. For example, the decoupling level metric, proposed by Mo *et al.* [67], measures the suitability of a software project to decouple into smaller independent modules. The authors found that it is easier to locate bugs and changes in software with higher decoupling levels. For the class level, one of the most common sets of code metrics is the C&K [20] that consists of six different metrics, including depth of inheritance, number of children, and weighted methods per class. The performance of the C&K metrics in estimating software maintenance, however, has been debated. While some studies reported negative results [28, 31], positive results were reported too [10]. A popular concept that is associated with product metrics is known as code smells. Some common code smells are God (or Blob) class—a class with multiple responsibilities; Brain method—a large and less cohesive method; Shotgun Surgery—a change in a class induces changes in many other classes, etc [17]. Most studies confirm that code smells negatively impact software maintenance [50, 78, 80, 106, 120]. However, some studies have surprisingly found little to no impact [75, 104, 105].

For the method level, the popular code metrics include size [23], cyclomatic complexity [65], nested block depth [25], code readability [15, 82], fanout [81], maintainability index [25], halstead metrics [3], comment ratio [109], etc. Unlike the class-level code metrics, method-level code metrics were found to be useful in multiple studies. For example, Landman *et al.* [54] found that code complexity at the method level can not be explained just by size, which is common in class-level granularity [31]. This means complexity metrics such as McCabe are useful at the method-level granularity to understand maintenance effort. A similar observation was made by Chowdhury *et al.* [23] where seven method-level code metrics were evaluated. They found that *context is king* while evaluating the effectiveness of code metrics.

Process metrics, on the other hand, are generally extracted from the change history of a code component [68, 81] that includes the number of added lines, deleted lines, number of times a component was modified, number of distinct authors associated with the code component, number of modified branches (e.g., if-else), etc. Developer-centric metrics capture the history of a developer, or even the focus of a developer on a specific code component that can be captured by the "scattering of changes by developers" [26].

**Motivation for product metrics.** In this paper, we focus on predicting whether a method will become highly change-prone (i.e., a *monstrous* method). Our prediction is made at the time the method is created, as identifying potential issues early can significantly reduce maintenance costs [18]. This approach confines us to using only product metrics, or source code metrics since change history is not available at a method's inception. However, our models can be enhanced to integrate process and developer-centric metrics for scenarios where change history is also available.

## 3 METHODOLOGY

In this section, we discuss about project selection, change history collection, and change- and bug-proneness indicators. We also talk about our selection of code metrics, data preprocessing, and statistical tests that we used in this paper. To enhance readability, here we do not discuss the methods that are specific only to a particular research question. Instead, we discuss those methods when we answer the research questions in Section 4.

### 3.1 Project Selection

To investigate change-proneness at the method level, we selected open-source software projects from GitHub [3], a popular platform for mining software repositories in software engineering research [47, 48]. Prior studies typically employed one of two approaches: aggregated analysis [31, 81, 106] or individual project analysis [46, 90, 101, 121]. Aggregated analysis can be skewed by outliers [1], while individual project analysis may introduce selection bias [31, 84]. To address these issues, we opted for individual project analysis with a large and unbiased sample of 49 projects. This set was compiled by taking the union of projects used in five different research studies [31, 35, 79, 88, 106]. We believe this extensive dataset will enable us to draw more generalizable conclusions, at least within the context of open-source software. Our study begins with an analysis of 774,051 source code methods, as detailed in Table 1.

[3]https://github.com/ last accessed: June-23-2024

### 3.2 Change History Collection

The nature of our research—predicting methods that are prone to significant changes—necessitates the collection of accurate change histories for numerous methods. While some previous studies have explored similar areas, they primarily relied on heuristic-based methods for constructing these histories. For instance, Spadini *et al.* [106] considered that if two methods in successive commits exhibit a cosine similarity of over 95%, they can be considered the same method. However, this heuristic lacks rigorous empirical validation.

Grund *et al.* [35] developed CodeShovel, a tool validated against a human-created oracle of 200 source code methods. CodeShovel utilizes the Jaro-Winkler distance algorithm to identify similar methods. When a method is renamed, CodeShovel first searches for similar methods within the same file, using threshold values derived from a training process. If no similar method is found, it assumes the method may have been moved to a different file. In this case, it parses all modified Java files, extracts the methods, and identifies the most similar one based on another set of threshold values established through training. If no method exceeds a defined similarity threshold, CodeShovel assumes it has located the introduction commit of that particular method. CodeShovel was also evaluated with independent industry developers. CodeShovel provides detailed information on when, why, and how a method was modified in a `git` repository.

To improve CodeShovel's accuracy even further, Jodavi et al. [42] introduced CodeTracker. Although CodeTracker demonstrated greater accuracy than CodeShovel, its evaluation was based on a modified version of the CodeShovel oracle and was not validated with independent industry developers. Additionally, CodeTracker's execution time was slower than that of CodeShovel.

To collect the change history of Java methods used in this study, we employed CodeShovel due to its faster execution time and rigorous evaluation with industry developers. We utilized the Javaparser library [4] to extract source code methods from the 49 GitHub projects. Our custom program then interfaced with CodeShovel to gather comprehensive change histories for these methods. CodeShovel has been effectively used in recent research requiring method history tracking [1, 23–25].

### 3.3 Selecting the Change-proneness Indicators

To capture the change-proneness of a method, we consider four different indicators: the number of revisions, the sum of all the diff sizes, the sum of all the numbers of added lines, and the sum of the edit distances.

**#Revisions.** The number of revisions of a code component has been widely used as an indicator of change-proneness in previous studies [5, 6, 68, 70, 81, 101]. The underlying assumption is that a well-designed code component should require fewer revisions. We determined the number of revisions for a method based on the total change commits reported by CodeShovel, excluding commits where no actual code modifications were made. This distinction is important because CodeShovel also reports commits where a method is moved or its container file is renamed, even if the method's content remains unchanged. In our publicly shared dataset, the column *DiffSizes* contains all the change sizes of a method in different change commits. Normally, the number of values (separated by commas) in that column would indicate the number of revisions for a particular method. However, we discarded those counts where the change size is 0. A zero value in a particular commit indicates that a method was moved without any content change, or its containing file was renamed in that commit—therefore, such a commit should not be considered as a change commit.

[4] https://github.com/javaparser/javaparser last accessed: June-24-2024

Table 1. The selected 49 projects to conduct our study. We have collected a total of 774,051 methods. The snapshot SHAs are presented as well to reproduce our results. For most of the projects, the number of contributors and stars is high, indicating the popularity of these projects to the open-source community.

| Repository | # Methods | #Contributors | #Stars | Snapshot |
|---|---|---|---|---|
| hadoop | 70,081 | 592 | 14,500 | 4c5cd7 |
| elasticsearch | 62,190 | 1,932 | 68,400 | 92be38 |
| flink | 38,081 | 908 | 16,764 | 261e72 |
| lucene-solr | 37,133 | 234 | 4,400 | b457c2 |
| presto | 36,715 | 734 | 15,700 | bb20eb |
| docx4j | 36,514 | 37 | 1,620 | 36c378 |
| hbase | 36,274 | 463 | 5,200 | 3bd542 |
| intellij-community | 35,950 | 1,035 | 16,800 | cdf2ef |
| weka | 35,639 | 2 | 323 | a22631 |
| hazelcast | 35,265 | 340 | 6,000 | a59ad4 |
| spring-framework | 26,634 | 544 | 43,641 | 1984cf |
| hibernate-orm | 24,800 | 420 | 4,678 | 2c12ca |
| eclipseJdt | 22,124 | 133 | 146 | 475591 |
| guava | 20,757 | 264 | 41,775 | e35207 |
| sonarqube | 20,627 | 137 | 5937 | 6b806e |
| jclouds | 20,358 | 231 | 384 | 7af4d8 |
| wildfly | 19,665 | 343 | 2,593 | f21f5d |
| netty | 16,908 | 526 | 27,185 | 662e0b |
| cassandra | 15,953 | 447 | 8,600 | 7cdad3 |
| argouml | 12,755 | 4 | 238 | fcbe6c |
| jetty | 10,645 | 159 | 3160 | fc5dd8 |
| voldemort | 10,601 | 65 | 2,475 | a7dbde |
| spring-boot | 10,374 | 828 | 56,378 | 199cea |
| wicket | 10,058 | 83 | 570 | e3f370 |
| ant | 9,781 | 58 | 303 | 1ce1cc |
| jgit | 9,548 | 140 | 983 | 855842 |
| mongo-java-driver | 9,467 | 151 | 2,422 | 8ab109 |
| pmd | 8,992 | 229 | 3482 | d115ca |
| xerces2-j | 8,153 | 3 | 38 | cf0c51 |
| RxJava | 8,145 | 278 | 44,922 | 880eed |
| openmrs-core | 6,066 | 371 | 1,017 | c5928a |
| javaparser | 5,862 | 157 | 3,771 | 8f25c4 |
| hibernate-search | 5,345 | 58 | 390 | 5b7780 |
| titan | 4,590 | 34 | 5,135 | ee226e |
| facebook-android-sdk | 3,759 | 124 | 6100 | fb1b91 |
| checkstyle | 3,340 | 287 | 6115 | 164a75 |
| commons-lang | 2,948 | 160 | 2,138 | f69235 |
| lombok | 2,684 | 114 | 10,391 | 4fdcdd |
| atmosphere | 2,659 | 111 | 3,524 | fadfb0 |
| jna | 2,636 | 143 | 6,655 | b8443b |
| Essentials | 2,390 | 215 | 1,086 | d36d80 |
| junit5 | 2,085 | 167 | 4,691 | be2aa2 |
| hector | 1,958 | 71 | 648 | a302e6 |
| okhttp | 1,953 | 236 | 40,428 | 5224f3 |
| mockito | 1,498 | 21 | 12,043 | 077562 |
| cucumber-jvm | 1,146 | 226 | 2,291 | b57b92 |
| commons-io | 1,145 | 76 | 787 | 11f0ab |
| vraptor4 | 926 | 48 | 343 | 593ce9 |
| junit4 | 874 | 151 | 8,160 | 50a285 |
| **Total** | **774,051** | | | |

**DiffSize and AdditionOnly.** The number of revisions alone may not fully capture a component's change-proneness. A single revision might involve minimal changes, such as modifying a few lines of code or comments, while another revision could involve substantial modifications, impacting maintenance costs and effort differently. Additionally, developers' commit habits can affect the number of revisions [115]. Therefore, diff size is often used as an alternative measure of change-proneness [68, 95, 101]. It is, however, argued that adding lines of code is generally more challenging than deleting lines, making the number of added lines only (AdditionOnly) a potentially more accurate indicator of change-proneness [68, 81, 101] than DiffSize.

**EditDistance.** A limitation of DiffSize and AdditionOnly measurements is that they do not account for the varying lengths of modified lines. EditDistance addresses this by quantifying the number of characters that need to be added, deleted, or replaced to transform one version of a source code into another. Consequently, we also used the Levenshtein edit distance [57, 92, 95, 107] as an indicator of change-proneness. To compute the EditDistance for a method, we summed the edit distances across all of its change commits. DiffSize and AdditionOnly were calculated in a similar manner.

### 3.4 Selection and Measurement of Code Metrics

From our literature review [23, 25, 29, 68, 81, 102] and domain knowledge about method-level code metrics, we have identified a set of 17 code metrics for our analysis. These metrics are presented in Table 2. These metrics cover the most important aspects of code maintainability, such as `size`, `testability`, `readability`, `dependency`, and `others`. `Size` is considered the most important code metric by many previous studies [25, 28, 31]. We calculated size as the source lines of code without comment and blank lines, consistent with practices in other research [25, 54, 86]. `Testability` assesses how easily a method can be tested, typically captured by cyclomatic complexity metrics like McCabe's [65], which counts the number of independent paths through a method. A higher McCabe score suggests that more effort and time are required for testing. However, McCabe's metric does not account for the number of control variables and comparisons within predicates, which is why we included these additional metrics in our study. `Readability` is another important indicator of maintenance effort [14, 44, 93, 94], as practitioners spend a significant amount of time in reading code. We included two different scores of readability: one according to Buse *et al.* [15] and the other according to Posnett *et al.* [82]. We considered the fanout measurements as an indication of dependency because if there is a problem in any of the called methods, that problem can propagate to the caller method. We also added the MaintainablityIndex because it is a composite metric that is adopted by some popular tools including the Verifysoft technology[5] and Visual Studio[6]. The other metrics used in our study are the number of local variables, number of parameters, commentToCode ratio, Halstead lengh, IndentSTD, isPublic, isStatic, and getter/setter.

[5]https://verifysoft.com/en_maintainability.html: last accessed: June-24-2024
[6]https://docs.microsoft.com/en-us/visualstudio/code-quality/code-metrics-maintainability-index-range-and-meaning?view=vs-2022: last accessed: June-24-2024

Table 2. Description of the 17 method-level code metrics.

| Metric | Description |
|---|---|
| **Size** | Source lines of code without comment and blank lines [25, 54] |
| **McCabe** | Cyclomatic complexity, calculated as 1 + #Predicates [65] |
| **NVAR** | Number of control variables that influence the decisions in conditionals [66] |
| **NCOPM** | Number of comparisons that influence the decisions in conditionals [66] |
| **IndentSTD** | A simple proxy of complexity metrics [23, 39] |
| **MaximumBlockDepth** | Maximum of the nested depth of all control structures [81] |
| **FanOut** | Number of called methods [81] |
| **Length** | Halstead length [3] |
| **MaintainabilityIndex** | A composite metric that leverages McCabe, Halstead volume, and size [23] |
| **Readability** | The readability metric developed by Buse *et al.* [15] |
| **SimpleReadability** | The readability metric developed by Posnett *et al.* [82] |
| **Parameters** | Total number of parameters in a method [81] |
| **Variables** | Total number of local variables inside a method |
| **CommentRatio** | Number of comment lines / Size [81] |
| **Getter/Setter** | Whether a method is a getter or a setter [24] |
| **isPublic** | Whethere a method is public or not |
| **isStatic** | Whether a method is static or not |

The aforementioned code metrics were collected from the source code of a method's introduction commit, and traced using CodeShovel. Our objective is to predict whether a method is *monstrous* (i.e., highly change-prone) immediately after it is added to the system, allowing for early optimizations. To measure these metrics, we extended our own tool previously used in several studies [1, 23–25]. This tool calculates various code metrics from a method's source code. It employs the JavaParser[7] library to parse the source code and generate an abstract syntax tree (AST), which serves as the primary source for metric calculations. The accuracy of this tool was rigorously validated by two researchers, who tested it on 500 randomly selected Java methods.

### 3.5 Statistical Tests, Graphs, and Age Normalization

In this paper, we calculate correlation coefficients between different distributions multiple times. Upon evaluating the data with the Anderson-Darling normality test [89], we found that most of our data do not follow a normal distribution. Consequently, we opted for the non-parametric Kendall's $\tau$, which is widely used in data-driven software engineering research [22, 31, 41]. Unless otherwise stated, the correlation coefficients reported are statistically significant ($P \leq 0.05$). For graphical representation, we predominantly use the cumulative distribution function (CDF) rather than the probability density function (PDF). The CDF's monotonic nature simplifies data interpretation, as it avoids the zigzag pattern often observed with the PDF.

In our dataset, the methods vary significantly in age, as they were introduced to the software projects at different times. This variation poses a challenge for our analysis because older methods have had more time to undergo changes compared to newly introduced methods, irrespective of their code metrics (e.g., size, McCabe complexity, readability). Figure 1 (a) illustrates that a method's age is positively correlated with the number of revisions it undergoes. Specifically, for approximately 80% of the projects, the correlation coefficients exceed 0.2. This suggests that without normalizing for age, our analysis could be inaccurate. To address this issue, we excluded methods that are less than five years old from our analyses. We chose a five-year threshold as it provides sufficient time to assess which methods are highly change-prone (i.e., *monstrous*). However, we cannot directly compare methods that are five years old with those that are more than five years old. Therefore, we also excluded any revisions made after the first five years of a method's lifetime. As shown in Figure 1 (b), this five-year threshold allows us to retain information from approximately 84% of the revisions, as most revisions in our dataset occurred within the first five years of a method's life. Unfortunately, applying this threshold resulted in the loss of four projects (rxJava, Titan, JUnit5, and Essentials) and approximately 43% of the methods in our dataset. Despite this, we prioritized retaining change information over the number of methods. We still retained data from 45 popular software projects with 407,202 source methods, which is still significantly higher than many other similar studies (e.g., [29, 30, 51, 53, 68, 81], to name a few).

[7] https://github.com/javaparser/javaparser last accessed: June-24-2024

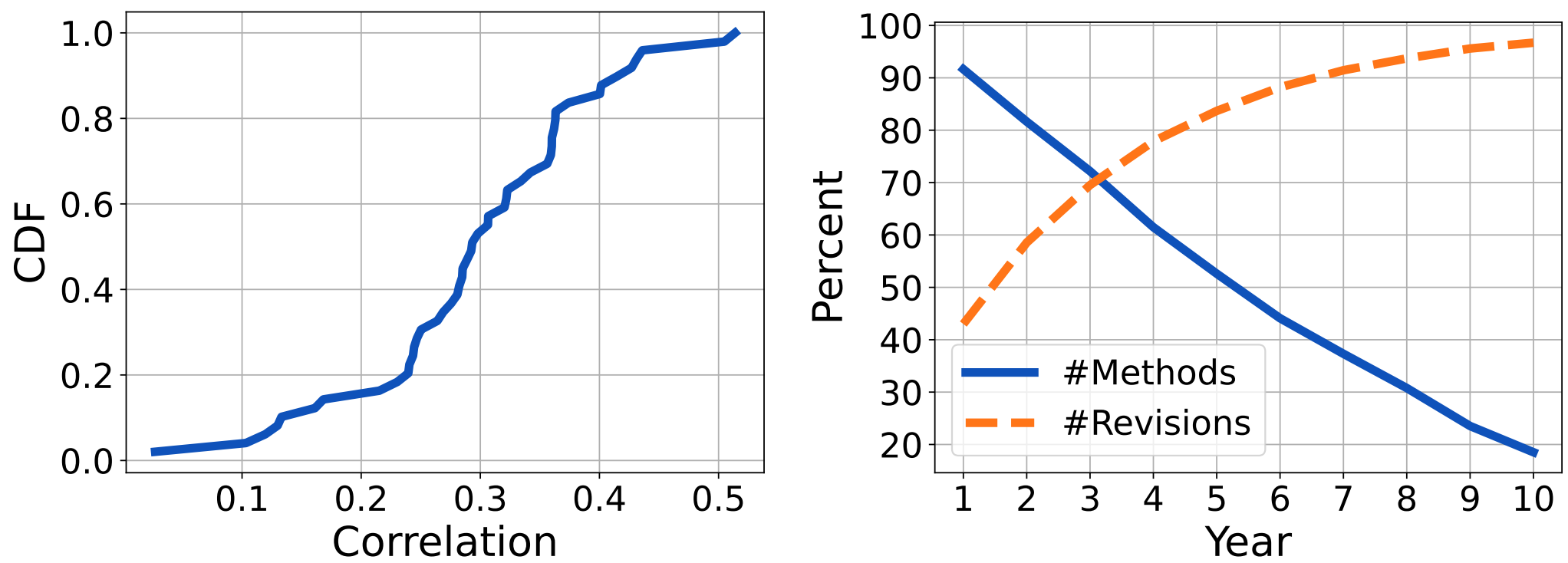

(a) Age vs. #Revisions

(b) Impact of different age threshold

Fig. 1. Figure (a) shows the CDF of correlations between methods' ages and their number of revisions for all 49 projects. Figure (b) depicts how different age thresholds impact the captured number of methods and number of revisions. The #Methods shows how many methods we lose with the increasing age threshold, whereas the #Revisions shows how many revisions we can retain with the increasing age threshold—with a higher age threshold, we retain more revision information but lose more methods.

## 4 ANALYSIS AND RESULTS

In this section, we discuss the findings based on our four research questions. Research question-specific methods are also discussed here.

### 4.1 RQ1: Evaluating the Pareto 80/20 Principle

Previous studies [35, 68, 99] have hinted that code churn is not uniformly distributed among all the code components (e.g., modules, files/classes, and methods). Based on this recommendation, we hypothesize that only a small percentage of source code methods are probably responsible for most code changes. Figure 2 shows the distributions (after ranking methods by their change-proneness) of the four change-proneness indicators for five randomly selected projects. For the #Revisions (Figure (a)), the distributions resemble long-tailed distributions [61]. This suggests that capturing only the top change-prone methods might miss a significant portion of the revisions across the entire dataset. This observation, however, changes significantly when we focus on change sizes instead of revisions, which is most noticeable when edit distance is considered (Figure (d)). The distributions no longer have long tails, indicating that the topmost methods are responsible for most changes.

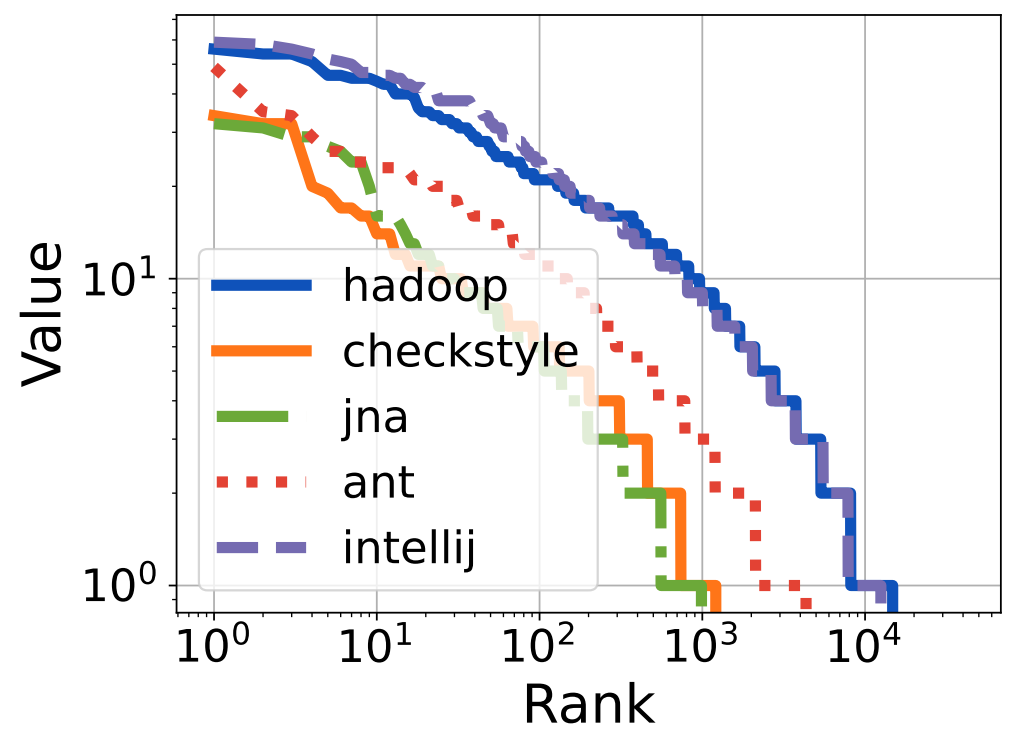

(a) #Revisions

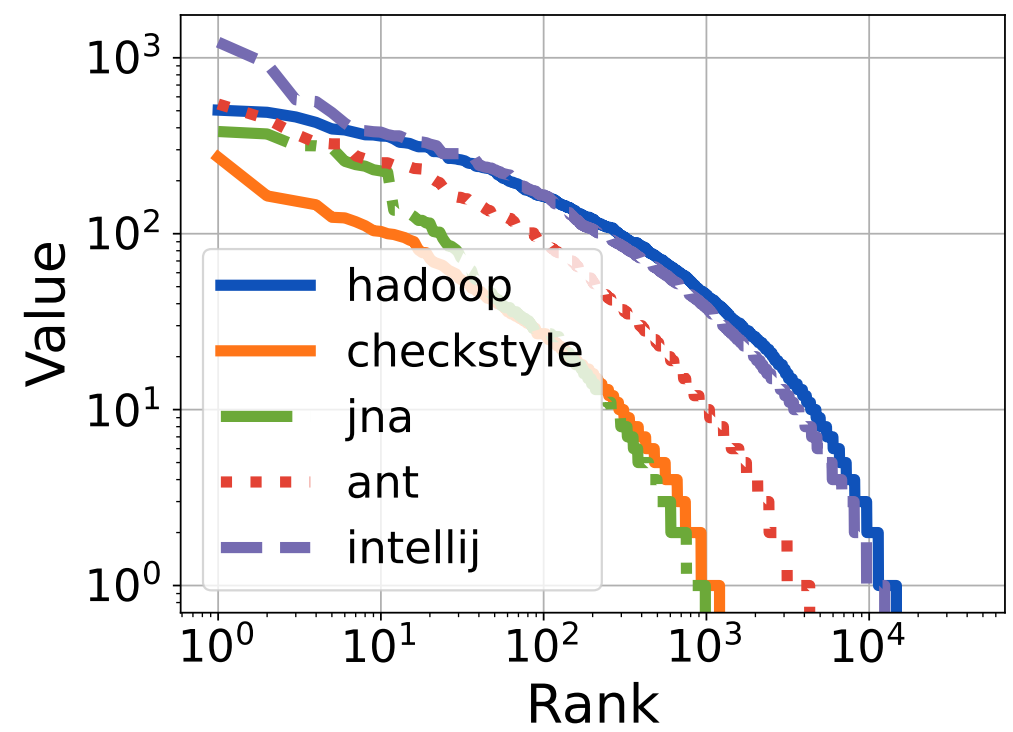

(b) AdditionOnly

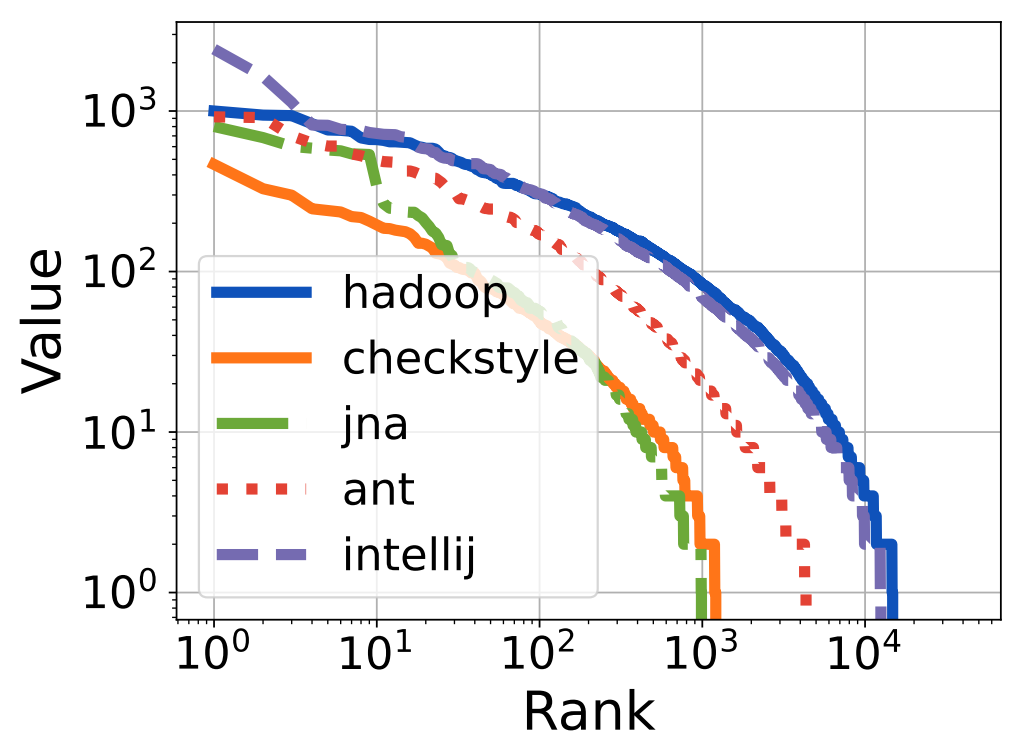

(c) Diffsize

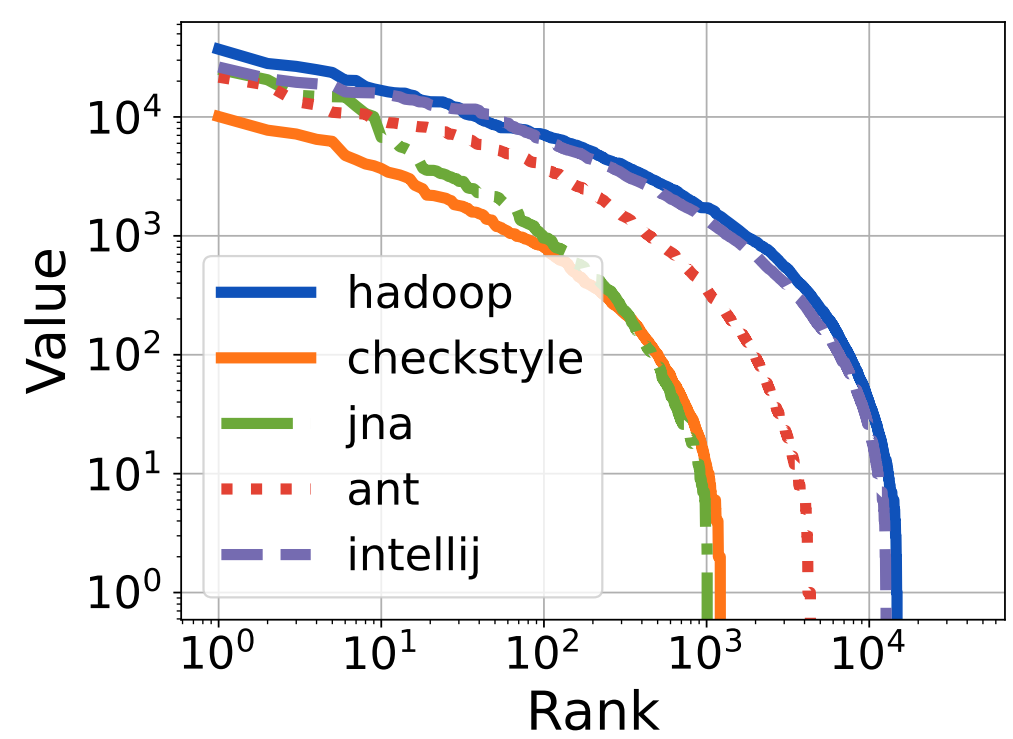

(d) EditDistance

Fig. 2. Log-log plots of the distributions of the four change-proneness indicators after ranking the source code methods for five randomly selected projects. Results are similar for all projects and are not included to enhance graph readability.

We further validate this observation using the widely known 80/20 Pareto principle [33]: *Do the top 20% of change-prone methods account for 80% of all changes in a software project?* If our hypothesis is true, it would mean that predictive models should focus on capturing these highly change-prone 20% methods (*monstrous* methods) to optimize resource allocation and reduce future maintenance costs and effort. Figure 3 displays the results for all four change-proneness indicators. For instance, Figure 3 (a) shows that if we rank methods based on the number of revisions and capture only the top 5% of such methods, for ≈80% of the projects, we can capture ≥30% of all the revisions that happened in a project. However, if we capture the top 20% of the methods, we can capture ≥60% of all the revisions that happened in a project. If we consider the edit distance as the change-proneness indicator, the results are even more encouraging. For most of the projects, we can capture more than 80% of all changes by capturing only the top 20% of the methods. In fact, for 80% of the projects, the change information that we can capture is ≥ 90%. These are the methods we call *monstrous*, and these are the methods we are interested in predicting at once they are pushed to a system.

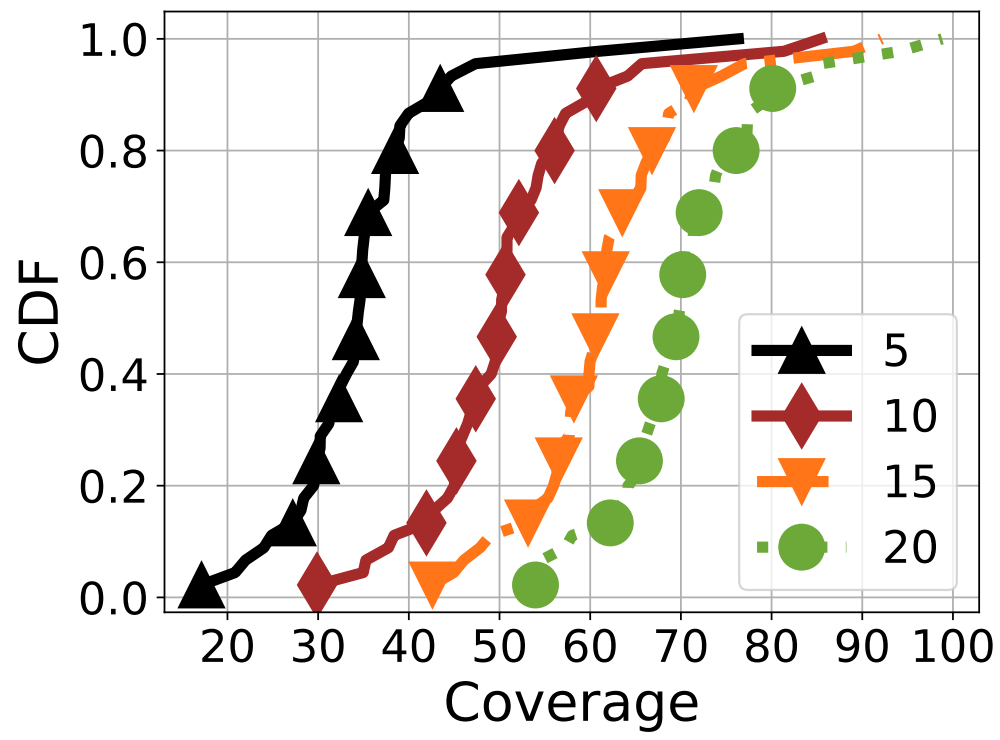

(a) #Revisions

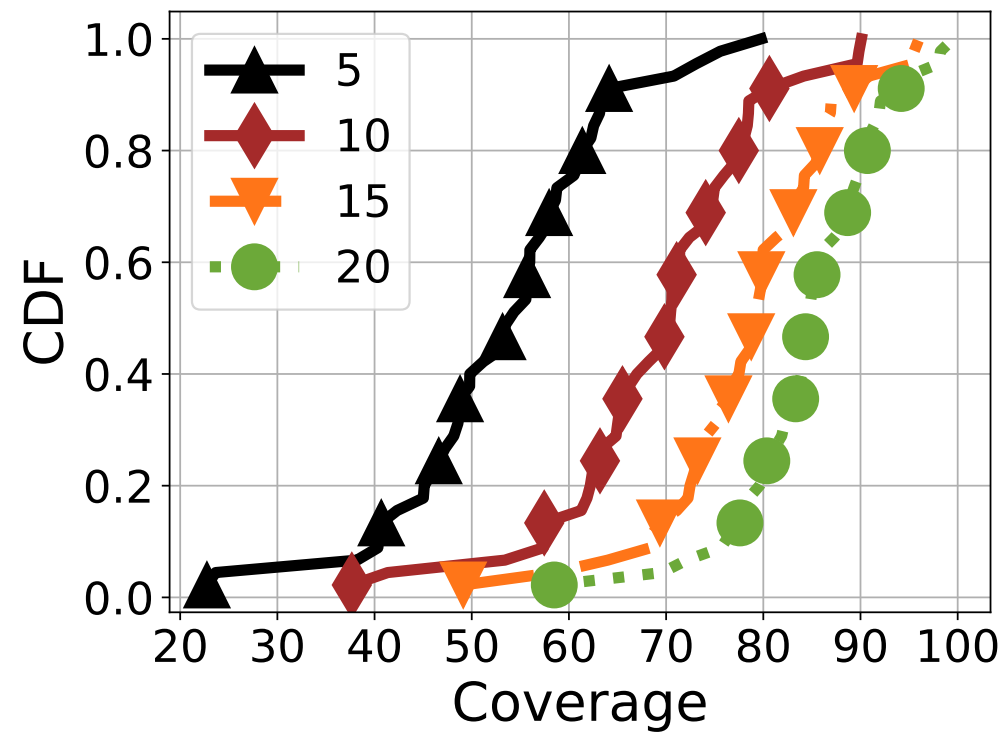

(b) AdditionOnly

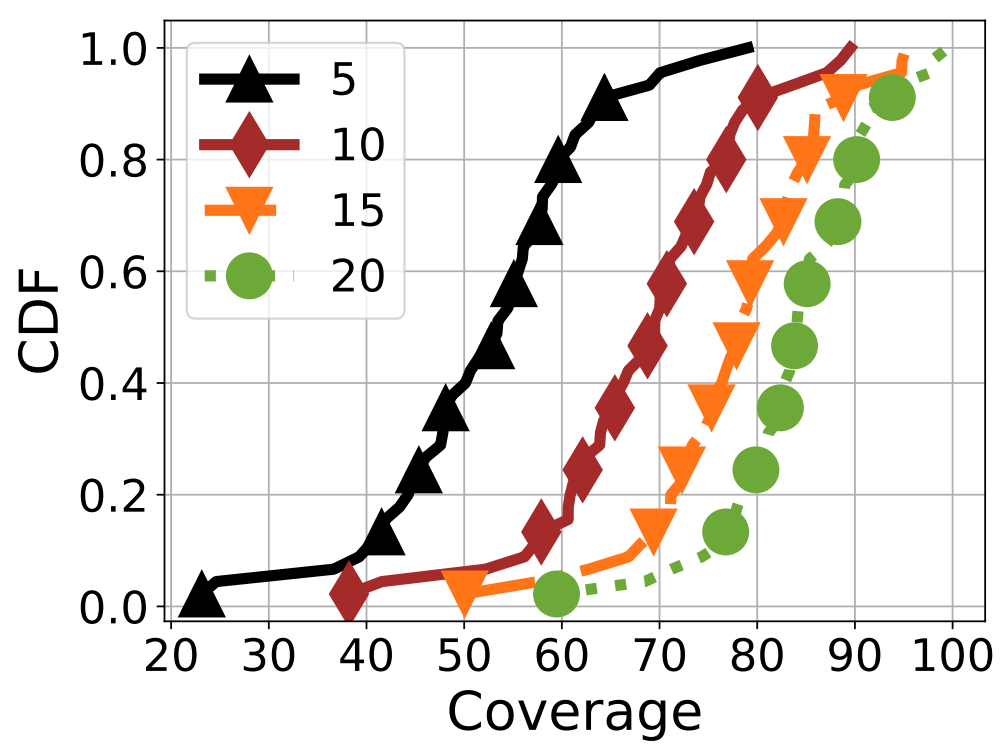

(c) Diffsize

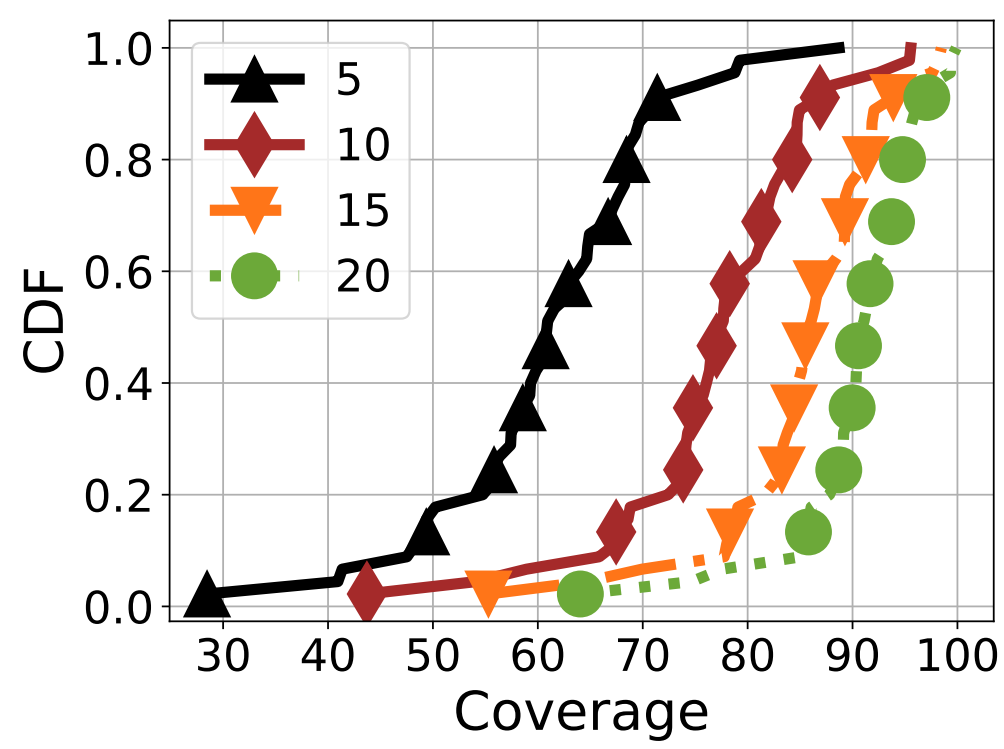

(d) EditDistance

Fig. 3. Evaluating the 80/20 rule (Pareto principle). In addition to the top 20% change-prone methods, we also evaluate the percent of changes that can be captured by capturing only the top 5%, 10%, and 15% change-prone methods. To enhance graph readability, less number of markers are used although each line contains the results of all 45 projects.

**Summary:** When using edit distance or other change-proneness indicators (excluding revisions), our analysis indicates that changes in open-source software projects adhere to a distribution similar to the Pareto 80/20 principle. This is encouraging because it means that by focusing on the top 20% change-prone methods in a project, we can capture the majority of the changes that occur within that project.

### 4.2 RQ2: Percent of Bugs Captured by the *Monstrous* Methods

Bug prediction is often referred to as *the prince of empirical software engineering research* [55], and yet bug prediction research did not have a significant impact on practitioners [55, 58, 114]. One of the main problems in building an accurate bug prediction model is the lack of accurately labeled data, often due to the complexity or irrelevance of changes included in a single commit [24, 36, 37]. To address this issue, we explore whether focusing on the most change-prone source code methods can also help identify the most bug-prone methods. In other words, we investigate if it is possible to capture a significant number of bug-prone methods without directly building and relying on a traditional bug prediction model. Given the challenges associated with accurate bug labeling, we created two different datasets for this investigation: one aimed at high recall and the other at high precision. We hope that, together, these datasets will provide more robust and reliable insights compared to those observations derived from a single dataset alone.

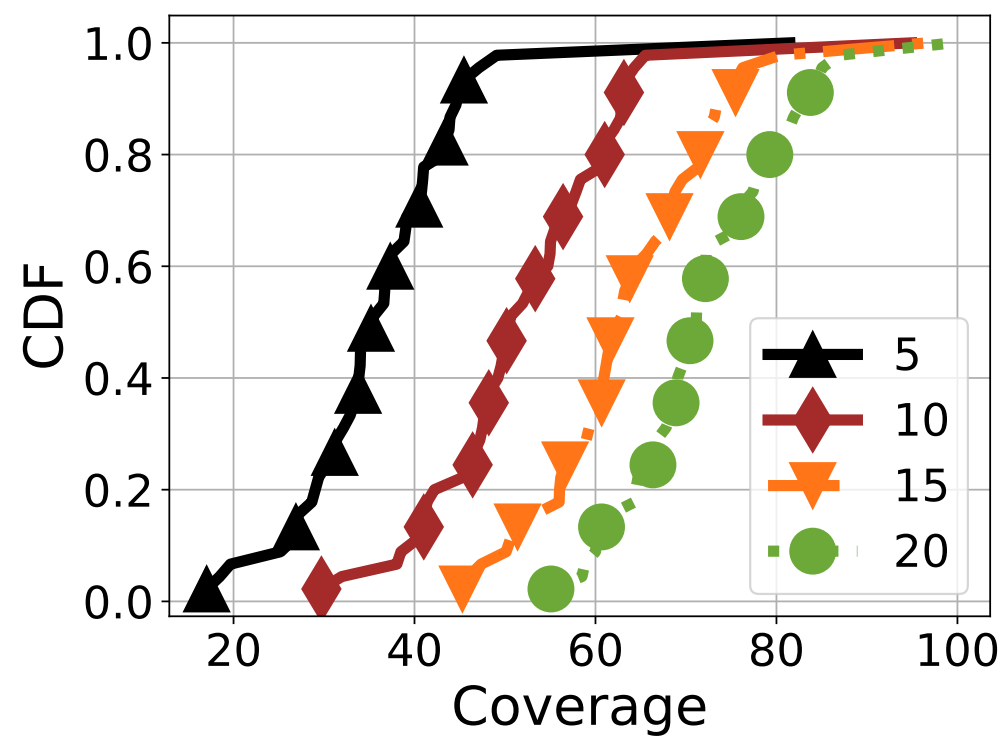

(a) High recall dataset

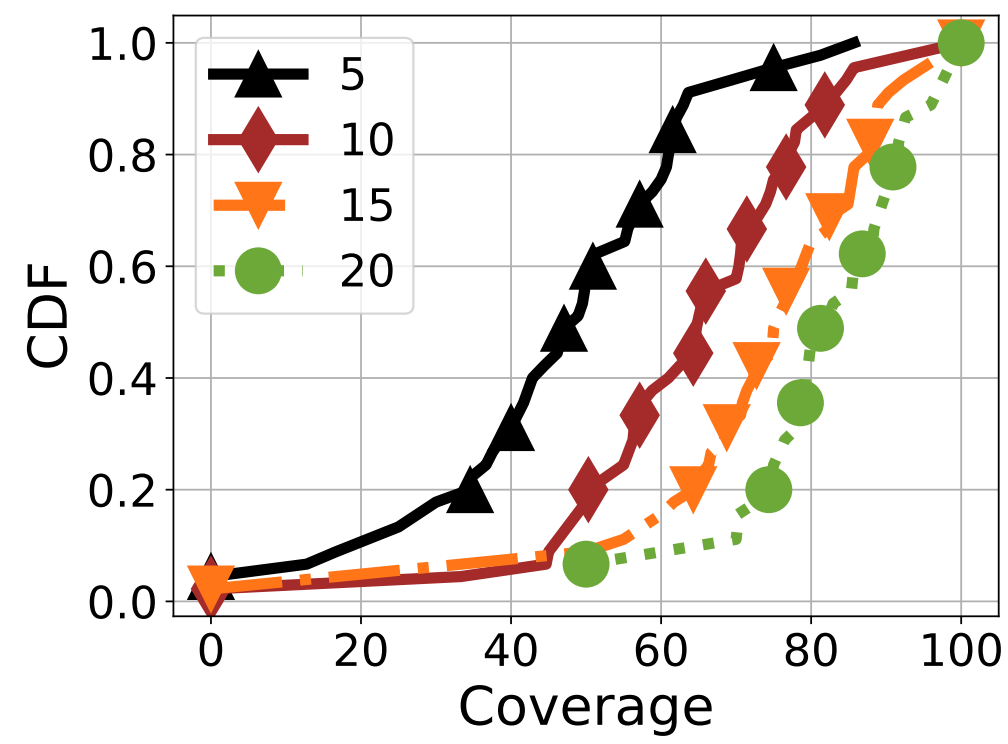

(b) High precision dataset

Fig. 4. Percent of total bugs in a project that can be captured by capturing the top 5%, 10%, 15%, and 20% change-prone methods only. Edit distance is used to represent change-proneness. Figure (a) and (b) show the results for the high recall and high precision datasets, respectively. The observations are similar if we use other change-proneness indicators.

- **High recall dataset.** We identified buggy methods by considering all methods associated with bug-fix commits. A commit was categorized as a bug-fix commit if its message included any of the following nine bug-related keywords (or their variants): *error*, *bug*, *fix*, *issue*, *mistake*, *incorrect*, *fault*, *defect*, and *flaw*. This keyword list has been utilized in several previous studies [69, 88]. While this approach is effective in identifying most buggy methods (high recall), it tends to generate numerous false positives due to the complications of tangled code changes.
- **High precision dataset.** To reduce the number of false positives while accepting a trade-off in the recall, Chowdhury *et a.* [24] proposed a more conservative approach to bug labeling. First, from the list of nine words, they removed the word *issue* as it produces too many false positives, and added the word *misfeature* based on the recommendation of Rosa *et al.* [91]. Additionally, they found that using both bug-related and fix-related words (e.g., *fix*, *address*, *resolve*) enhances the precision of identifying bug-fix commits. To further reduce the impact of tangled changes, they recommended excluding bug-fix commits that modify more than one method. Their rationale is that a commit affecting only a single method can be labeled as buggy with high accuracy. However, this approach inevitably reduces recall, as it results in the exclusion of many potential bug-fix commits.

Figure 4 (a) illustrates the results for the high recall dataset. It shows that focusing on the top 5% of change-prone methods captures approximately 40% of the total bugs, but this is observed in only about 40% of the projects. In contrast, targeting the top 20% of highly change-prone methods allows for capturing over 60% of the bugs across approximately 80% of the projects. The results are even more promising for the high precision dataset, as depicted in Figure 4 (b). Here, capturing the top 20% of highly change-prone methods enables the detection of more than 80% of the total bugs in around 80% of the projects.

**Summary:** Identifying the *monstrous* methods—the top 20% of highly change-prone methods—not only aids in predicting the majority of future changes a project will undergo but also helps in detecting most software bugs. This is particularly encouraging as it shows an alternative approach to bug detection, potentially bypassing the need for complex and often impractical bug prediction models.

### 4.3 RQ3: Predicting the *Monstrous* Methods

In our previous analyses, we established the significance of identifying the *monstrous* source code methods to reduce software maintenance efforts and costs. We now explore whether these methods can be predicted at their inception. Specifically, we examine if a single code metric can effectively identify the *monstrous* methods within a software project. Using a single metric for this purpose would greatly benefit practitioners by making the process of locating and optimizing problematic code much faster and more efficient. To investigate this, we selected two widely used method-level code metrics: Size and McCabe. Below, we explain the rationale for selecting these two metrics.

- **Size** is widely recognized as one of the most effective code metrics [54, 68, 81]. Some studies have even suggested that size is the only useful code metric due to its strong correlation with other metrics [28, 31, 105]. Indeed, size has been described as *the code metric that rules them all* [25]. We argue that if only one metric were to be selected for understanding code maintenance, size would be the most logical choice.
- **McCabe**, also known as cyclomatic complexity, is likely to be the second most popular code metric in the method-level source code granularity. It has been employed in nearly all studies of method-level maintenance that utilized static code metrics (e.g., [24, 29, 30, 68, 81, 102], to name a few). Unlike coarser-granularity metrics [31], McCabe remains valuable at the method level even when accounting for the confounding effects of size [23, 54].

Figure 5 shows the distributions of the correlation coefficients for Kendall's $\tau$. Considering all the change-proneness indicators, the correlations between size and change-proneness are weak (within 0.3 [87]) for 20% of the projects, and moderate (within 0.7 [87]) for the remaining projects. Even for a single project, no strong or perfect correlation was observed. When McCabe is used in place of size, the correlations are either similar to or lower than those observed with size. This observation does not change if we use any of the other code metrics. To enhance the readability of the paper, we share all other results with our replication package (`extra-results/rq3/`). We conclude that a single code metric is insufficient for accurately identifying the *monstrous* methods.

The limitations of using a single code metric to identify the *monstrous* methods prompted us to explore more complex models that leverage multiple features. We selected five machine learning algorithms to develop and evaluate model accuracy: Logistic Regression, Decision Tree, Random Forest, AdaBoost, and Neural Networks. These algorithms were chosen to represent a diverse range of machine-learning techniques. **Random Forest**, an ensemble learning algorithm, was an obvious choice due to its widespread use and strong performance in similar classification studies [29, 68, 81]. In contrast, **AdaBoost** is another ensemble method that emphasizes boosting rather than bagging, as employed by Random Forest. Random Forest's approach to aggregate decisions from multiple decision trees leads to increased training times. Therefore, we aimed to assess the efficiency of a single **Decision Tree** as an alternative. We also included **Logistic Regression** as a straightforward function-based learning algorithm alongside the more complex **Neural Networks**. This selection allows us to explore the performance differences across both simple and advanced machine-learning techniques. To construct and evaluate these models, we employed two different approaches that ensure no historical data is needed to test a project. This approach ensures that the models can be applied early in the software development lifecycle, even when no historical data is available.

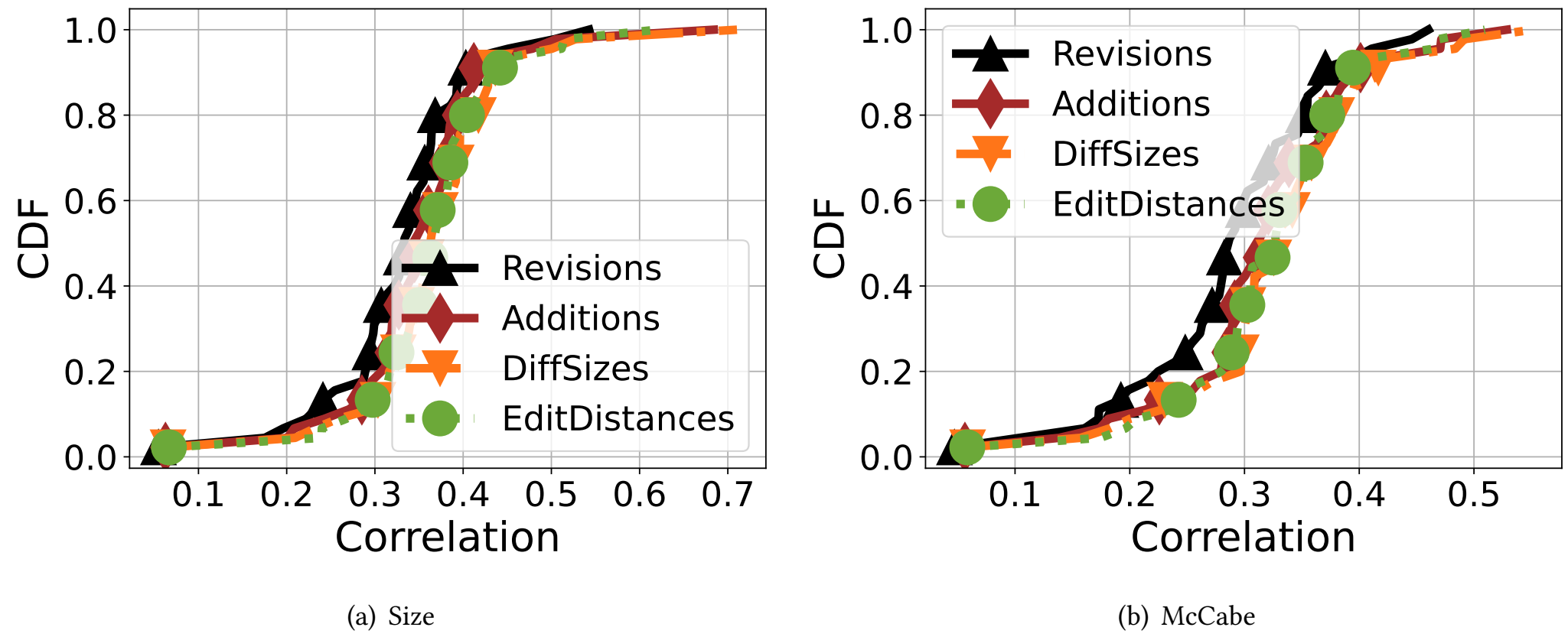

Fig. 5. The correlation coefficients between size and change-proneness (a) and McCabe and change-proneness (b) for the 45 software projects.

- **Approach 1.** We randomly divided the projects into three sets: 20% for testing, 10% for validation, and 70% for training. We tuned the model parameters (e.g., strength of regularization) using only the training and validation sets. Subsequently, we evaluated the accuracy of our models on the test set.
- **Approach 2.** We employed the leave-one-out approach [117], where each project was used as the test set once while the remaining projects were used for training. This process was repeated 45 times, allowing us to test each of the 45 projects individually. This approach helps assess whether the model accuracies vary significantly across different projects.

To model our data, we classified the methods into three categories: *monstrous* (the top 20% most change-prone methods), *good* (methods that did not change in their first five years, as discussed in Section 3), and *bad* (methods that are neither *monstrous* nor *good*). In our subsequent analysis, we present results using edit distance as the change-proneness indicator, as it provides a more accurate measure of maintenance effort by accounting for character-level differences between two source code versions of a method [92, 95, 107].

Initially, we observed poor prediction performance across all machine learning algorithms. Upon manual inspection, we discovered that methods in the *bad* category often exhibited characteristics similar to either *good* or *monstrous* methods, introducing noise into the dataset and hindering robust model training. For instance, some methods, which had small size and low complexity similar to the *good* methods, were unexpectedly changed once (e.g., a comment modification) and thus categorized as *bad* instead of *good*. Conversely, some methods, which were changed only a few times, were inaccurately labeled as *monstrous* because they fell within the top 20% most change-prone methods, due to the generally low number of revisions across those projects. This noise in the dataset hindered the robust training and the development of accurate prediction models.

One potential solution to reduce the noise in data labeling is to merge all the *good* and *bad* methods into one category. Alternatively, a simpler approach is to discard all the *bad* methods from the dataset and to reformulate the prediction task as a binary classification problem (*monstrous* vs. *good*), rather than a multiclass classification problem. We adopted the latter approach, discarding all *bad* methods (approximately 36% of methods per project on average) from our dataset. This approach is still practical for practitioners who aim to emulate *good* methods and avoid *monstrous* ones. However, the resulting dataset was imbalanced, with significantly fewer *monstrous* methods compared to *good* methods, which impacted the robustness of model training. To address this imbalance, we used the *RandomOverSampler* from the `sklearn` library[8] to oversample the *monstrous* methods in the training data. This oversampling was applied only to the training set, while the validation and test sets remained unaltered.

Table 3. Results for **Approach 1**, rounded to 2 decimal places. Nine randomly selected projects were used for testing: *hector*, *okhttp*, *mockito*, *pmd*, *hazelcast*, *vraptor4*, *javaparser*, *openmrs-core*, and *jetty*. We used the sklearn Python library to run these algorithms. The best sets of parameters for these algorithms were selected with ad-hoc search. We began with the default parameter values and refined them by experimenting with various random values close to the defaults. We stopped when we found that no more improvement in accuracy (e.g., F-measure) was achievable after some changes to the parameters. We acknowledge that our results could be improved with additional time and effort devoted to finding better parameters. The high AUC score implies that the models are good at discriminating *monstrous* methods from *good* methods. The Matthews Correlation Coefficients (MCC) indicate that our predictions are much better than random prediction (MCC is ≈0 for average random prediction).

| Algorithm | Precision | Recall | F-measure | AUC | MCC |
|---|---|---|---|---|---|
| **LogisticRegression** | 0.77 | 0.79 | 0.78 | 0.83 | 0.65 |
| **DecisionTree** | 0.74 | 0.81 | 0.78 | 0.83 | 0.65 |
| **RandomForest** | 0.75 | 0.81 | 0.78 | 0.83 | 0.66 |
| **Adaboost** | 0.76 | 0.80 | 0.78 | 0.83 | 0.65 |
| **NN** | 0.77 | 0.80 | 0.78 | 0.83 | 0.66 |

Table 3 shows the results with **Approach 1**. The precision and recall values are based on targeting the *monstrous* methods—in `sklearn`, `pos_label='monstrous'` was set. If we do the same calculation targeting the *good* methods, all the scores improve significantly. For example, the F-measure becomes 0.87 with the DecisionTree algorithm. This implies that predicting the *monstrous* methods is more difficult than predicting the *good* methods, partly due to the fewer number of *monstrous* methods than the *good* methods. Overall, all machine learning algorithms demonstrated similar performance and achieved good prediction accuracy. To ensure thoroughness, we also assessed the models without addressing the data imbalance issue, maintaining the training set as is. The results are presented in Table 4. Notably, the precision of all models improved. This is expected—with a significantly fewer number of *monstrous* methods in the dataset, the models would be more conservative when predicting a method as *monstrous*. However, the imbalance in the dataset severely affected recall, significantly impacting the overall F-score. Practitioners seeking high precision in predicting *monstrous* methods may consider using models trained on the imbalanced dataset, but this approach comes at the cost of considerable recall sacrifice.

We also examined whether excluding highly correlated features would enhance the models' performance. Given that code metrics are generally highly correlated with *size* [25, 28, 31], we calculated Kendall's $\tau$ correlation coefficients between size and other code metrics. All code metrics exhibited correlation coefficients of $\geq 0.5$ (or $\leq -0.5$) with size, except for *CommentRatio*, *Readability*, *IndentSTD*, *isPublic*, *isStatic*, *Getter/Setter*, and *Parameters*. Consequently, we constructed models using only size and these selected metrics. The results are summarized in Table 5. Interestingly, we can observe a decline in performance in most cases. This outcome is not entirely surprising, as previous research has indicated that correlated features can enhance model performance [43], particularly when the training dataset is sufficiently large. Additionally, contrary to findings from file- or class-level studies [31], it has been shown that code metrics remain valuable for understanding method-level maintenance effort, despite their high correlation with size [23, 54].

[8]https://scikit-learn.org/stable/ (last accessed: 28-Jun-2024)

Table 4. Results for **Approach 1**. Everything is similar to Table 3, except the data imbalance issue was not addressed during training. With the data imbalance issue, the precision of the models has improved, but the recall, F1 score, and AUC have deteriorated significantly.

| Algorithm | Precision | Recall | F-measure | AUC | MCC |
|---|---|---|---|---|---|
| **LogisticRegression** | 0.87 | 0.57 | 0.70 | 0.76 | 0.60 |
| **DecisionTree** | 0.86 | 0.56 | 0.68 | 0.76 | 0.58 |
| **RandomForest** | 0.84 | 0.62 | 0.71 | 0.78 | 0.60 |
| **Adaboost** | 0.84 | 0.64 | 0.73 | 0.79 | 0.62 |
| **NN** | 0.86 | 0.59 | 0.70 | 0.77 | 0.60 |

Table 5. Results for **Approach 1**. This time we have excluded code metrics that are highly correlated to *size*.

| Algorithm | Precision | Recall | F-measure | AUC | MCC |
|---|---|---|---|---|---|
| **LogisticRegression** | 0.72 | 0.74 | 0.73 | 0.80 | 0.58 |
| **DecisionTree** | 0.73 | 0.78 | 0.76 | 0.81 | 0.62 |
| **RandomForest** | 0.75 | 0.77 | 0.76 | 0.82 | 0.63 |
| **Adaboost** | 0.72 | 0.80 | 0.76 | 0.81 | 0.61 |
| **NN** | 0.76 | 0.76 | 0.76 | 0.82 | 0.63 |

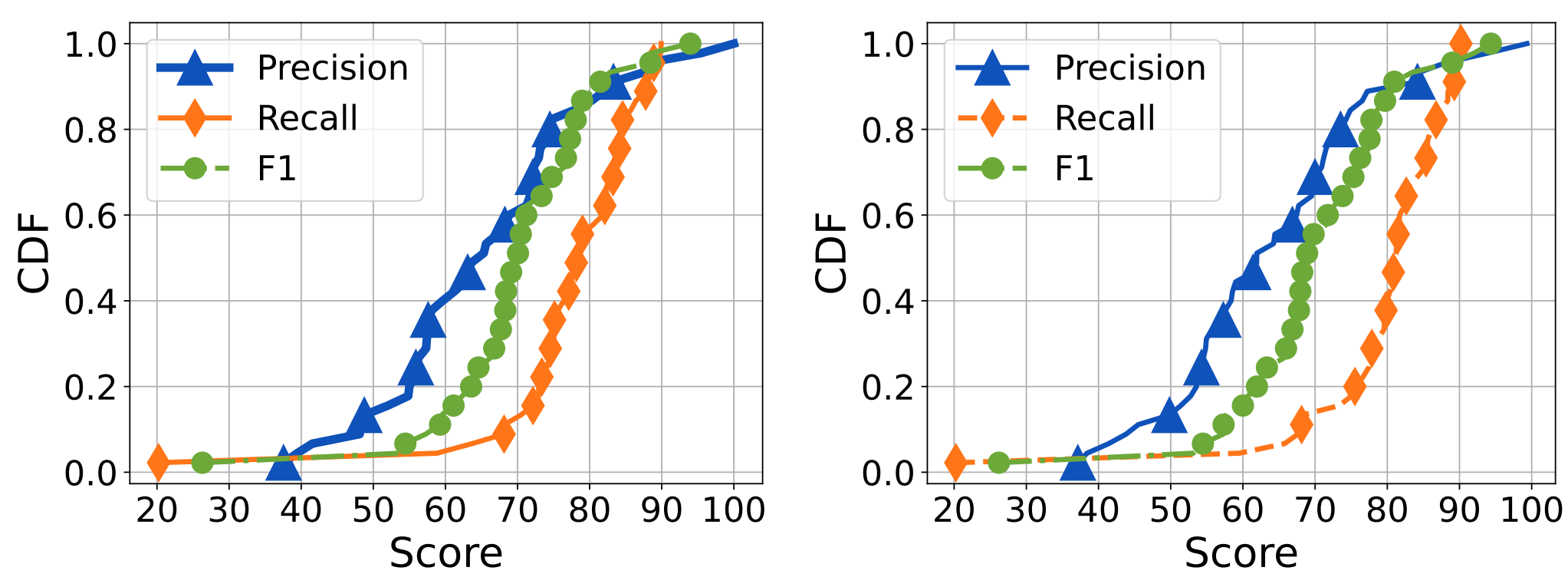

(a) LogisticRegression

(b) DecisionTree

Fig. 6. Results for the leave-one-out approach; in each of the 45 rotations, a different project was selected for testing, and others were selected for training. Results are shown only for the LogisticRegression and DecisionTree algorithms because of very similar observations with all others.

We also sought to understand whether the accuracies of our algorithms vary significantly across different projects. This was addressed using the leave-one-out approach (i.e., **Approach 2**). The results are illustrated in Figure 6. As shown, the accuracies for most projects are comparable to, or even higher than, those observed in Table 3. The high recall is promising, indicating that our predictive models can effectively capture most *monstrous* methods. However, for some projects, the prediction accuracies are notably lower. We hypothesize that these projects exhibit unique characteristics and may require model building tailored to similar projects for better prediction accuracy. This is discussed in Section 5 as a potential future work.

**Summary:** Capturing *monstrous* methods with a single code metric (such as size or McCabe) proves challenging. However, machine learning models generally perform well in predicting these *monstrous* methods. These models can be effectively used to identify methods that are likely to undergo significant future changes (and, consequently, generate bugs). Nevertheless, the prediction accuracies can vary depending on the specific projects being tested.

### 4.4 RQ4: Investigating the Cases Where Prediction is Difficult.

We want to know when machine learning models fail to capture the monstrous methods. To be more precise, we want to investigate the cases where a method with characteristics similar to the *monstrous* methods (e.g., large, complex, and less readable) is not *monstrous*, and vice versa. To facilitate this analysis, we first ranked the source methods based on a composite score, similar to a study by Mashhadi *et al.* [64]. The composite score for each method was derived from Kendall's $\tau$ correlation coefficients, which were calculated between the 17 code metrics (as detailed in Table 2) and the edit distances across all projects. Table 6 shows the results.

The correlation coefficients are negative for five code metrics: *Readability*, *SimpleReadability*, *MaintainabilityIndex*, *isPublic*, and *Getter/Setter*. Except for the *isPublic* metric, the negative correlation between change-proneness and the other four metrics is unsurprising. A code that has a higher *readability*, and *MaintainabilityIndex* score should be less change-prone. Similarly, a getter or setter should not change much as they are generally small in size and do very simple predefined tasks. Among the 12 metrics with positive correlations to change-proneness, the *CommentRatio* stands out as surprising. We anticipated a negative correlation because a higher *CommentRatio* should enhance readability and reduce maintenance effort. However, previous studies [21, 63] have also found a positive correlation between *commentRatio* and maintenance effort because comments are generally found in complex code.

To rank the methods based on their code metric values, we first normalized each metric using a 0—1 scaling approach. This normalization ensures that for each project, all feature values are between 0 and 1, allowing features with different value ranges to have comparable weights in the aggregation process. For metrics with negative correlation coefficients with change-proneness (as shown in Table 6), we inverted the values by multiplying them by 1. We then computed the composite score for each method by summing all normalized feature values. Metrics with binary values, such as *isPublic*, *isStatic*, and *Getter/Setter*, were excluded from this calculation due to their limited informational value in this context. We then sorted methods within each project based on their composite scores. From this ranking, we identified two groups: the top 50 methods, labeled as *surprisingly good*, which did not change despite their high composite scores (e.g., large size, high McCabe complexity, and low readability score); and the bottom 50 methods, labeled as *surprisingly monstrous*, which changed frequently despite their low composite scores (e.g., small size, low McCabe complexity, and high readability score). To ensure generalizability, we limited the maximum number of selected methods per project to 4 only: 2 for *surprisingly good* and 2 for *surprisingly monstrous* methods.

To label these 100 methods with meaningful concepts, we followed an open coding approach [112] which is extensively used in grounded theory [40, 111]. We developed the concepts inductively, creating new ones only when a method did not fit any previously established categories. To label each method, we examined its source code, comments, change history, commit messages, and, when relevant, other co-evolved methods. This process makes it difficult to have a larger sample size. However, 100 samples are considered as the minimum acceptable number of samples for qualitative analysis [76, 77].

Table 6. Correlation coefficients between the 17 method-level code metrics and edit distances (aggregated data). All the coefficients are statistically significant ($P \leq 0.05$), indicating a reliable relationship between the metrics and edit distances. The directions of the correlation coefficients (positive or negative) are the same if we analyze each project individually. None of the correlation coefficients are strong, confirming our previous finding that a single code metric cannot accurately predict a method's change-proneness.

| Metric | Correlation |
|---|---|
| **Size** | 0.34 |
| **McCabe** | 0.30 |
| **NVAR** | 0.28 |
| **NCOMP** | 0.28 |
| **IndentSTD** | 0.17 |
| **MaximumBlockDepth** | 0.31 |
| **FanOut** | 0.33 |
| **Length** | 0.34 |
| **MaintainabilityIndex** | -0.33 |
| **Readability** | -0.25 |
| **SimpleReadability** | -0.31 |
| **Parameters** | 0.18 |
| **Variables** | 0.32 |
| **CommentRatio** | 0.19 |
| **Getter/Setter** | -0.23 |
| **isPublic** | -0.12 |
| **isStatic** | 0.04 |

To enhance the internal validity of our labeling approach, we enlisted a graduate research student in software engineering who has a decade of industry experience, primarily in Java. The student was briefed on the methodology used to collect the history of the methods and was tasked with independently labeling the selected 100 methods. Following this initial labeling, we facilitated an in-person discussion to merge similar labels and address any disagreements. For instance, we combined specific labels like "sorting" and "searching" into a broader category labeled as *Algorithm*. We encountered 5 and 8 instances of disagreement for the *surprisingly good* and *surprisingly monstrous* methods, respectively. To resolve these disagreements, we engaged another graduate student in software engineering with two years of industry experience. All disagreements were ultimately resolved through a majority vote. Table 7 shows the identified concepts of the 50 *surprisingly monstrous* and the 50 *surprisingly good* methods.

*4.4.1 Surprisingly monstrous methods.* These methods initially had low composite scores (i.e., small, less complex, and highly readable), which led the machine learning models to inaccurately predict them as *good*. We identified five common concepts that help explain why these methods evolved into *monstrous* rather than remaining *good*.

- **Incomplete.** Some of these methods were initially incomplete or had blank implementations. Over time, some methods evolved due to the adoption of Test-Driven Development (TDD)[11], where each commit introduced a new testing scenario. Additionally, some methods started with sub-optimal implementations and self-admitted technical debts [83]. For example, the method `onConnect()` (`11870.json`) from the `jetty` project was initially blank and included a self-admitted technical debt (to-do). It has since undergone significant evolution.
- **NonCriticalChanges.** Many of these methods underwent significant changes in edit distances due to modifications in comments or formatting. For instance, the method `getToplevelAST` (`2636.json`) from the `checkstyle` project experienced substantial comment changes. Recent research indicates that code metrics exhibit greater predictive power when such non-critical changes are discarded [1].

Table 7. The identified concepts from the *surprisingly monstrous* and *surprisingly good* methods. The *surprisingly monstrous* methods have undergone numerous changes, despite their similarity to the *good* methods, which are characterized by small, less complex code, high readability scores, etc. Examples of these methods are available in the `rq5_samples/surprisingly_monstrous` directory of our dataset. In contrast, the *surprisingly good* methods have remained unchanged, even though they resemble the *monstrous* methods, which are characterized by large, complex code, low readability scores, etc. Examples of these methods can be found in the `rq5_samples/surprisingly_good` directory.

| Surprisingly monstrous | | | Surprisingly good | | |
|---|---|---|---|---|---|
| **Concept** | **Count** | **Example** | **Concept** | **Count** | **Example** |
| **Incomplete** | 20 | 11870.json | **Setups** | 18 | 8097.json |
| **NonCriticalChanges** | 11 | 2636.json | **Algorithms** | 15 | 76134.json |
| **PossibleCodeShovelError** | 10 | 17764.json | **InaccurateLabels** | 7 | 5910.json |
| **NewScenario/TrialandError** | 5 | 17655.json | **Parsing** | 5 | 24018.json |
| **Other** | 4 | 32988.json | **Other** | 5 | 1843.json |

- **PossibleCodeShovelError.** We discovered that CodeShovel's accuracy diminishes when tracking the change history of small methods, which is problematic because *surprisingly monstrous* methods are generally very small. CodeShovel can be confused by two different small methods that have similar string similarities. Out of 50 such methods, CodeShovel inaccurately traced the history of 10 methods. Our manual analysis revealed that these 9 methods did not change significantly but were mislabeled as *monstrous*. For instance, the `close` method (`17764.json`) in the `flink` project was incorrectly linked to the `clear` method from a different file, thus inaccurately capturing the change history of the `clear` method.
- **NewScenario/TrialandError.** Despite being very small and less complex, these methods have evolved significantly due to new scenarios that developers initially overlooked. In some cases, uncertainty about whether a particular implementation approach would work led developers to experiment with multiple solutions, resulting in extensive change histories and large changes. A notable example is the `free` method (`17655.json`) from the `netty` project.
- **Other.** These methods were altered for seemingly random reasons, making them challenging to conceptualize and predict.

We propose that future research could leverage these concepts to enhance the prediction accuracy of *monstrous* methods. For instance, machine learning techniques can be used to identify methods with technical debt [119]. Additionally, ignoring NonCriticalChanges could improve the classification of methods as *good* or *monstrous*. We are actively working on refining CodeShovel to better track very small methods. However, methods categorized as NewScenario/TrialandError and Other present a challenge due to the difficulty in understanding their nature early on. Encouragingly, these methods constitute only about 18% of the *surprisingly monstrous* methods.

*4.4.2 Surprisingly good methods.* These methods were large, complex, and not readable, which led the machine learning models to predict them as *monstrous*. However, these methods never changed in the first five years of their lifetime. We found five common concepts that can be used to explain these methods—why they are *good* instead of being *monstrous* (highly change-prone).

- **Setups.** These methods implement some straightforward setups or configurations. They often have many `if-else` or `switch-cases` to check conditions and configure/set accordingly. For example, the method `endElement` (`8097.json`) in the `docx4j` project configures font, height, etc., for a UI component. Despite their large size and high complexity, these methods change rarely.
- **Algorithms.** These methods implement some algorithms, such as sorting, matching, and even machine learning-based classification. An example is the `mainSimpleSort` method (`76134.json`) of the `hadoop` project. Once implemented and committed, these methods do not evolve.
- **Inaccurate.** Although our ranking approach found very high composite scores for these methods, these methods are not very similar to the other *monstrous* methods and should not be considered *monstrous*. For example, the `geoWithinBox` method (`5910.json`) of the `mongo-java-driver` project is small in size and has a very small McCabe score, but it was still ranked with a high composite score. This was due to lots of tabs inside the code, making the `IndentSTD` score very high which impacted the composite score very significantly. This is common in the other 6 methods in this category. If we knew this, we could have discarded the `IndentSTD` metric while calculating the composite scores. `IndentSTD` score is supposed to replace McCabe-like metric [39], but apparently, that is not the case for some methods.
- **Parsing.** These methods parse input sources to extract specific data of interest. The `parse` method (`24018.json`) of the `lucene-solr` project is an example of this concept.
- **Other.** These methods are unique and do not fit into a common theme or concept.

Similar to the *surprisingly monstrous* methods, most of the *surprisingly good* methods can be conceptualized. If we can develop models to detect these common concepts in large and complex source code methods, we can use these concepts as features to build more accurate prediction models. Machine-learning models will learn that the likelihood of a large and complex source method being highly change-prone is reduced if it fits one of the identified concepts.

**Summary:** The surprising methods become less surprising once their true nature is understood. Many instances where source code metrics fail to predict a method's change-proneness can be attributed to one of the frequently occurring concepts we identified. Future research could focus on developing models to detect these concepts and integrate them as features to improve the prediction accuracy of change prediction models.

## 5 DISCUSSION

In this paper, we discussed the importance of predicting source code change-proneness to help reduce future software maintenance efforts and costs. Unlike other previous studies, we focused on the method-level code granularity, which is often preferred by practitioners and researchers [35, 68, 81]. We found that change-proneness at the method level often follows the Pareto 80/20 principle (**RQ1**). In addition, these small number of methods account for most of the identified bugs in the studied open-source software projects (**RQ2**). These observations are encouraging: detecting only a small number of methods in their early lifetime can help reduce software maintenance significantly. Although detecting these highly change-prone methods using a single code metric is challenging, machine learning models that utilize multiple metrics prove effective in their detection (**RQ3**). Our manual analysis reveals that methods that are difficult to predict can be categorized into common themes or concepts (**RQ4**).

### 5.1 Implications for Researchers

Our findings confirm that internal source code metrics are valuable for predicting future maintenance burdens (e.g., by building accurate change prediction models). This is particularly encouraging, as it challenges previous studies that questioned the utility of code metrics in software maintenance research [28, 31, 105]. Consequently, researchers should prioritize the development of additional code metrics to enhance maintenance prediction models. Furthermore, based on the positive outcome, our results advocate for a greater emphasis on method-level maintenance models, which are more favored by software practitioners [35, 81], as opposed to class- or file-level models. Below, we also outline some potential future works that can improve future change-prediction models.

Future research should leverage our identified themes in *surprisingly monstrous* and *surprisingly good* methods to build new feature sets to further improve the prediction accuracy of the difficult-to-predict source methods. With the leave-one-out approach, we observed that the prediction accuracy is not similar across the studied projects. Some outlier projects showed low accuracy while prediction was performed on them separately. Future studies can focus on this aspect by developing models for similar project selection. The rationale is that a model should perform poorly if it is trained on a set of projects that are significantly different than the test project. Previous work—although in different contexts such as bug prediction—has shown the promise of selecting similar projects during model training [8, 24, 113, 116].

The accuracy of our models can potentially be improved by employing a mixture of models. Rather than relying on a single generic model, researchers could develop separate models tailored to different method sizes. Additionally, using an ensemble of machine learning models—where multiple base models contribute to the final prediction through majority voting—could improve prediction accuracy [73]. Furthermore, investigating *concept drift* could be useful. For instance, research could explore whether two similar methods introduced at different stages of the same project (e.g., when the project was new versus when it was mature) exhibit similar change-proneness. Previous studies have demonstrated that *concept drift* can significantly affect the performance of maintenance prediction models [9, 13, 27, 45, 116].

Future research should explore additional source code metrics, such as the *fanin* metric. A high *fanin* value, indicating that many other methods call a particular method, suggests that developers may be less inclined to modify it due to the widespread impact of such changes on dependent methods. Although we recognized the importance of this metric, we were unable to measure it due to the complexity of constructing thousands of call graphs for each project individually. This challenge arises from our need to predict a method's change-proneness at its introduction, requiring us to collect code metrics at the commit when the method was added to the software project. For instance, in the `hadoop` project alone, preprocessing and building call graphs approximately 27,000 times would be necessary, corresponding to its 27,000 commits.

In this study, we chose not to include developer-centric information to ensure our approach is applicable from the start of a software project. At the project's beginning, there is often limited history regarding the contributors specific to that project. Future research could explore how the histories of a project's contributors from other projects might be leveraged.

### 5.2 Implications for Practitioners

Practitioners can utilize our models to identify the *monstrous* methods as soon as they are integrated into a software system. By concentrating their maintenance efforts on a small subset of critical methods, they can optimize resource allocation and reduce maintenance costs. Early identification of high-risk methods would enable practitioners to implement proactive strategies that address potential issues before they escalate, thereby minimizing future maintenance burdens. Our findings indicate that the *monstrous* methods are not only responsible for 80% of future changes but also responsible for the majority of bugs, allowing software teams to streamline their bug-fixing processes. This focused approach will not only improve overall software quality but also reduce the time spent on debugging. The use of internal source code metrics can also be used by code reviewers to assess specific attributes of methods, such as complexity and readability. This targeted evaluation can provide deeper insights into the maintainability of the code.

### 5.3 Threats to Validity

Several threats may have impacted the validity of our findings.

***Construct validity*** is hampered by our selection of the CodeShovel tool for constructing method change history. As we have found in this study, CodeShovel's accuracy dwindles for small-size methods. CodeShovel, however, outperformed other tools from both research and practice, including FinerGit and IntelliJ / git log [35]. Another recent method change history tracer, CodeTracker [42], can be used in the future to evaluate our results. We considered the number of revisions and different change sizes as change-proneness indicators. Other change-proneness indicators, where the type of changes is also considered [1], should also be considered in the future. To detect bug-proneness, we relied on the keyword-based approach which is known to suffer from low precision. We mitigated this threat by constructing two different bug-proneness datasets: high precision and high recall datasets. In *RQ4*, we manually inspected and labeled all the concepts for the selected *surprisingly monstrous* and *surprisingly good* methods. This process is inherently susceptible to human bias and error. To mitigate this threat, we validated the identified concepts with the help of an undergraduate research student.

***Internal validity*** can be questioned due to our selection of certain statistical tests such as Kendall's $\tau$ correlation coefficient. Kendall's $\tau$, however, is commonly used in data-driven software engineering research [23, 31, 41]. Also, our selection of the hyperparameters was not systematic: we started with the default values and tried to improve the accuracy with some random trials. More systematic approaches (e.g., sequential model-based optimization [59]) can help improve our results.

***External validity*** is hampered due to our reliance on Java-based open-source software projects. We do not know if our results are applicable to closed-source software projects and projects implemented in different programming languages than Java.

***Conclusion validity*** of our study is affected by all the aforementioned threats. In addition, to normalize the impact of a method's age, we used five years of age as the threshold: we discarded methods that were younger than five years and considered changes that happened within the first five years only. We do not know if other thresholds, such as 10 years, would impact the conclusion of our study.

## 6 CONCLUSION

In this paper, we demonstrated that a significant majority of code changes (approximately 80%) and bugs are concentrated in a small subset of methods (around 20%). By identifying and optimizing these high-impact methods as soon as they are added to a project repository, substantial reductions in future software maintenance efforts can be achieved. Our analysis reveals that machine learning models, leveraging commonly used static code metrics, can effectively identify and optimize these critical methods at their inception. Additionally, we have uncovered specific themes or concepts that explain why certain methods are challenging to predict. For instance, small and less complex methods may become highly change-prone due to factors such as technical debt or the use of test-driven development practices. We anticipate that these insights will offer valuable guidance to the software maintenance research community, facilitating the development of new features and enhancing future models for predicting the highly change- and bug-prone code components.

## 7 ACKNOWLEDGMENTS

Shaiful Chowdhury's research is supported by an NSERC Discovery Grant, a University of Manitoba Research Grant Program (URGP), and a University of Manitoba start-up research grant.

## REFERENCES

[1] Syed Ishtiaque Ahmad, Shaiful Chowdhury, and Reid Holmes. [n. d.]. Impact of Methodological Choices on the Analysis of Code Metrics And Maintenance. https://papers.ssrn.com/sol3/papers.cfm?abstract_id=4805882. *Under Revision, Journal of Systems and Software* ([n. d.]).

[2] Reem Aleithan. 2021. Explainable just-in-time bug prediction: Are we there yet?. In *2021 IEEE/ACM 43rd International Conference on Software Engineering: Companion Proceedings (ICSE-Companion)*. 129–131.

[3] M. Alfadel, A. Kobilica, and J. Hassine. 2017. Evaluation of Halstead and Cyclomatic Complexity Metrics in Measuring Defect Density. In *2017 9th IEEE-GCC Conference and Exhibition*. 1–9.

[4] H. Alsolai, M. Roper, and D. Nassar. 2018. Predicting Software Maintainability in Object-Oriented Systems Using Ensemble Techniques. In *2018 IEEE International Conference on Software Maintenance and Evolution*. 716–721.

[5] V. Antinyan, M. Staron, J. Derehag, M. Runsten, E. Wikström, W. Meding, A. Henriksson, and J. Hansson. 2015. Identifying complex functions: By investigating various aspects of code complexity. In *2015 Science and Information Conference (SAI)*. 879–888.

[6] V. Antinyan, M. Staron, W. Meding, P. Österström, E. Wikstrom, J. Wranker, A. Henriksson, and J. Hansson. 2014. Identifying risky areas of software code in Agile/Lean software development: An industrial experience report. In *IEEE Conference on Software Maintenance, Reengineering, and Reverse Engineering*. 154–163.

[7] Elvira-Maria Arvanitou, Apostolos Ampatzoglou, Alexander Chatzigeorgiou, and Paris Avgeriou. 2017. A method for assessing class change proneness. In *Proceedings of the 21st International Conference on Evaluation and Assessment in Software Engineering*. 186–195.

[8] Takuya Asano, Masateru Tsunoda, Koji Toda, Amjed Tahir, Kwabena Ebo Bennin, Keitaro Nakasai, Akito Monden, and Kenichi Matsumoto. 2021. Using Bandit Algorithms for Project Selection in Cross-Project Defect Prediction. In *2021 IEEE International Conference on Software Maintenance and Evolution (ICSME)*. 649–653.

[9] Abdul Ali Bangash, Hareem Sahar, Abram Hindle, and Karim Ali. 2020. On the Time-Based Conclusion Stability of Cross-Project Defect Prediction Models. *Empirical Softw. Engg.* 25, 6 (2020).

[10] V.R. Basili, L.C. Briand, and W.L. Melo. 1996. A validation of object-oriented design metrics as quality indicators. *IEEE Transactions on Software Engineering* 22, 10 (1996), 751–761.

[11] Kent Beck. 2003. *Test-driven development: by example.* Addison-Wesley Professional.

[12] Robert M. Bell, Thomas J. Ostrand, and Elaine J. Weyuker. 2011. Does Measuring Code Change Improve Fault Prediction?. In *Proceedings of the 7th International Conference on Predictive Models in Software Engineering* (Banff, Alberta, Canada) *(Promise '11)*. Article 2, 8 pages.

[13] Kwabena E Bennin, Nauman bin Ali, Jürgen Börstler, and Xiao Yu. 2020. Revisiting the Impact of Concept Drift on Just-in-Time Quality Assurance. In *2020 IEEE 20th International Conference on Software Quality, Reliability and Security (QRS)*. 53–59.

[14] J. Börstler and B. Paech. 2016. The Role of Method Chains and Comments in Software Readability and Comprehension—An Experiment. *IEEE Transactions on Software Engineering* 42, 9 (2016), 886–898.

[15] Raymond P. L. Buse and Westley R. Weimer. 2010. Learning a Metric for Code Readability. *IEEE Trans. Softw. Eng.* 36, 4 (July 2010), 546–558.

[16] Gemma Catolino, Fabio Palomba, Andrea De Lucia, Filomena Ferrucci, and Andy Zaidman. 2018. Enhancing change prediction models using developer-related factors. *Journal of Systems and Software* 143 (2018), 14–28.

[17] Gemma Catolino, Fabio Palomba, Francesca Arcelli Fontana, Andrea De Lucia, Andy Zaidman, and Filomena Ferrucci. 2020. Improving change prediction models with code smell-related information. *Empirical Software Engineering* 25 (2020), 49–95.

[18] Celerity. [n. d.]. The True Cost of a Software Bug: Part One. https://www.celerity.com/insights/the-true-cost-of-a-software-bug. [Online; last accessed 01-Sep-2022].

[19] Yaohui Chen, Peng Li, Jun Xu, Shengjian Guo, Rundong Zhou, Yulong Zhang, Tao Wei, and Long Lu. 2020. Savior: Towards bug-driven hybrid testing. In *2020 IEEE Symposium on Security and Privacy (SP)*. 1580–1596.

[20] S. R. Chidamber and C. F. Kemerer. 1994. A metrics suite for object oriented design. *IEEE Transactions on Software Engineering* 20, 6 (1994), 476–493.

[21] Istehad Chowdhury and Mohammad Zulkernine. 2010. Can complexity, coupling, and cohesion metrics be used as early indicators of vulnerabilities?. In *Proceedings of the 2010 ACM Symposium on Applied Computing*. 1963–1969.

[22] Shaiful Chowdhury, Stephanie Borle, Stephen Romansky, and Abram Hindle. 2019. GreenScaler: training software energy models with automatic test generation. *Empirical software engineering : an international journal* 24, 4 (2019), 1649–1692.

[23] Shaiful Chowdhury, Reid Holmes, Andy Zaidman, and Rick Kazman. 2022. Revisiting the Debate: Are Code Metrics Useful for Measuring Maintenance Effort? *Empirical Software Engineering (EMSE)* 27, 6 (2022), 31 pages.

[24] Shaiful Chowdhury, Gias Uddin, Hadi Hemmati, and Reid Holmes. 2024. Method-level Bug Prediction: Problems and Promises. *ACM Transactions on Software Engineering and Methodology* 33, 4 (apr 2024).

[25] Shaiful Chowdhury, Gias Uddin, and Reid Holmes. 2022. An Empirical Study on Maintainable Method Size in Java. In *Proceedings of the International Conference on Mining Software Repositories (MSR)*. 252–264.

[26] Dario Di Nucci, Fabio Palomba, Giuseppe De Rosa, Gabriele Bavota, Rocco Oliveto, and Andrea De Lucia. 2018. A Developer Centered Bug Prediction Model. *IEEE Transactions on Software Engineering* 44, 1 (2018), 5–24.

[27] Jayalath Ekanayake, Jonas Tappolet, Harald C Gall, and Abraham Bernstein. 2009. Tracking concept drift of software projects using defect prediction quality. In *2009 6th IEEE International Working Conference on Mining Software Repositories*. 51–60.

[28] K. El Emam, S. Benlarbi, N. Goel, and S. N. Rai. 2001. The confounding effect of class size on the validity of object-oriented metrics. *IEEE Transactions on Software Engineering* 27, 7 (2001), 630–650.

[29] Rudolf Ferenc, Péter Gyimesi, Gábor Gyimesi, Zoltán Tóth, and Tibor Gyimóthy. 2020. An automatically created novel bug dataset and its validation in bug prediction. *Journal of Systems and Software* 169 (2020).

[30] Emanuel Giger, Marco D'Ambros, Martin Pinzger, and Harald C. Gall. 2012. Method-level bug prediction. In *Proceedings of the 2012 ACM-IEEE International Symposium on Empirical Software Engineering and Measurement*. 171–180.

[31] Yossi Gil and Gal Lalouche. 2017. On the Correlation between Size and Metric Validity. *Empirical Software Engineering* 22, 5 (Oct. 2017), 2585–2611.

[32] T.L. Graves, A.F. Karr, J.S. Marron, and H. Siy. 2000. Predicting fault incidence using software change history. *IEEE Transactions on Software Engineering* 26, 7 (2000), 653–661.

[33] Abraham Grosfeld-Nir, Boaz Ronen, and Nir Kozlovsky. 2007. The Pareto managerial principle: when does it apply? *International Journal of Production Research* 45, 10 (2007), 2317–2325.

[34] Felix Grund, Shaiful Chowdhury, Nick C. Bradley, Braxton Hall, and Reid Holmes. 2021. CodeShovel: A Reusable and Available Tool for Extracting Source Code Histories. In *Proceedings of the International Conference on Software Engineering: Companion Proceedings (ICSE-Companion)*. 221–222.

[35] Felix Grund, Shaiful Chowdhury, Nick C. Bradley, Braxton Hall, and Reid Holmes. 2021. CodeShovel: Constructing Method-Level Source Code Histories. In *Proceedings of the International Conference on Software Engineering (ICSE)*. 1510–1522.

[36] Steffen Herbold, Alexander Trautsch, Benjamin Ledel, Alireza Aghamohammadi, Taher A Ghaleb, Kuljit Kaur Chahal, Tim Bossenmaier, Bhaveet Nagaria, Philip Makedonski, Matin Nili Ahmadabadi, et al. 2022. A fine-grained data set and analysis of tangling in bug fixing commits. *Empirical Software Engineering* 27, 6 (2022).

[37] Kim Herzig, Sascha Just, and Andreas Zeller. 2016. The impact of tangled code changes on defect prediction models. *Empirical Software Engineering* 21 (2016), 303–336.

[38] K. Herzig and A. Zeller. 2013. The impact of tangled code changes. In *2013 10th Working Conference on Mining Software Repositories*. 121–130.

[39] A. Hindle, M. W. Godfrey, and R. C. Holt. 2008. Reading Beside the Lines: Indentation as a Proxy for Complexity Metric. In *16th IEEE International Conference on Program Comprehension*. 133–142.

[40] Rashina Hoda. 2021. Socio-technical grounded theory for software engineering. *IEEE Transactions on Software Engineering* 48, 10 (2021), 3808–3832.

[41] Laura Inozemtseva and Reid Holmes. 2014. Coverage is Not Strongly Correlated with Test Suite Effectiveness. In *Proceedings of the International Conference on Software Engineering (ICSE)*. 435–445.

[42] Mehran Jodavi and Nikolaos Tsantalis. 2022. Accurate method and variable tracking in commit history. In *Proceedings of the 30th ACM Joint European Software Engineering Conference and Symposium on the Foundations of Software Engineering*. 183–195.

[43] Elsa Johansson Lannge. 2021. Does the removal of correlated variables affect the classification accuracy of machine learning algorithms?

[44] J. Johnson, S. Lubo, N. Yedla, J. Aponte, and B. Sharif. 2019. An Empirical Study Assessing Source Code Readability in Comprehension. In *2019 IEEE International Conference on Software Maintenance and Evolution*. 513–523.

[45] Md Alamgir Kabir, Jacky W Keung, Kwabena E Bennin, and Miao Zhang. 2019. Assessing the significant impact of concept drift in software defect prediction. In *2019 IEEE 43rd Annual Computer Software and Applications Conference (COMPSAC)*, Vol. 1. 53–58.

[46] D. Kafura and G. R. Reddy. 1987. The Use of Software Complexity Metrics in Software Maintenance. *IEEE Transactions on Software Engineering* SE-13, 3 (1987), 335–343.

[47] Eirini Kalliamvakou, Georgios Gousios, Kelly Blincoe, Leif Singer, Daniel M German, and Daniela Damian. 2014. The promises and perils of mining github. In *Proceedings of the 11th working conference on mining software repositories*. 92–101.

[48] Eirini Kalliamvakou, Georgios Gousios, Kelly Blincoe, Leif Singer, Daniel M German, and Daniela Damian. 2016. An in-depth study of the promises and perils of mining GitHub. *Empirical Software Engineering* 21 (2016), 2035–2071.

[49] Yasutaka Kamei and Emad Shihab. 2016. Defect Prediction: Accomplishments and Future Challenges. In *2016 IEEE 23rd International Conference on Software Analysis, Evolution, and Reengineering (SANER)*, Vol. 5. 33–45.

[50] Foutse Khomh, Massimiliano Di Penta, and Yann-Gael Gueheneuc. 2009. An exploratory study of the impact of code smells on software change-proneness. In *2009 16th Working Conference on Reverse Engineering*. 75–84.

[51] Foutse Khomh, Massimiliano D. Penta, Yann-Gaël Guéhéneuc, and Giuliano Antoniol. 2012. An exploratory study of the impact of antipatterns on class change- and fault-proneness. *Empirical software engineering : an international journal* 17, 3 (2012), 243–275.

[52] Hiroyuki Kirinuki, Yoshiki Higo, Keisuke Hotta, and Shinji Kusumoto. 2014. Hey! are you committing tangled changes?. In *Proceedings of the 22nd International Conference on Program Comprehension*. 262–265.

[53] A Güneş Koru and Hongfang Liu. 2007. Identifying and characterizing change-prone classes in two large-scale open-source products. *Journal of Systems and Software* 80, 1 (2007), 63–73.

[54] D. Landman, A. Serebrenik, and J. Vinju. 2014. Empirical Analysis of the Relationship between CC and SLOC in a Large Corpus of Java Methods. In *IEEE International Conference on Software Maintenance and Evolution*. 221–230.

[55] Michele Lanza, Andrea Mocci, and Luca Ponzanelli. 2016. The Tragedy of Defect Prediction, Prince of Empirical Software Engineering Research. *IEEE Software* 33, 6 (2016), 102–105.

[56] V. Lenarduzzi, A. Sillitti, and D. Taibi. 2017. Analyzing Forty Years of Software Maintenance Models. In *International Conference on Software Engineering Companion (ICSE-C)*. 146–148.

[57] Vladimir I Levenshtein. 1966. Binary codes capable of correcting deletions, insertions, and reversals. In *Soviet physics doklady*, Vol. 10. 707–710.

[58] Chris Lewis, Zhongpeng Lin, Caitlin Sadowski, Xiaoyan Zhu, Rong Ou, and E. James Whitehead. 2013. Does bug prediction support human developers? Findings from a Google case study. In *2013 35th International Conference on Software Engineering (ICSE)*. 372–381.

[59] Gang Luo. 2016. A review of automatic selection methods for machine learning algorithms and hyper-parameter values. *Network Modeling Analysis in Health Informatics and Bioinformatics* 5 (2016), 1–16.

[60] Alan MacCormack, John Rusnak, and Carliss Y Baldwin. 2006. Exploring the structure of complex software designs: An empirical study of open source and proprietary code. *Management Science* 52, 7 (2006), 1015–1030.

[61] Aniket Mahanti, Niklas Carlsson, Anirban Mahanti, Martin Arlitt, and Carey Williamson. 2013. A tale of the tails: Power-laws in internet measurements. *IEEE Network* 27, 1 (2013), 59–64.

[62] Ruchika Malhotra and Ravi Jangra. 2017. Prediction & assessment of change prone classes using statistical & machine learning techniques. *Journal of Information Processing Systems* 13, 4 (2017), 778–804.

[63] Haroon Malik, Istehad Chowdhury, Hsiao-Ming Tsou, Zhen Ming Jiang, and Ahmed E Hassan. 2008. Understanding the rationale for updating a function's comment. In *2008 IEEE International Conference on Software Maintenance*. IEEE, 167–176.

[64] Ehsan Mashhadi, Shaiful Chowdhury, Somayeh Modaberi, Hadi Hemmati, and Gias Uddin. 2024. An empirical study on bug severity estimation using source code metrics and static analysis. *Journal of Systems and Software* 217 (2024), 112179.

[65] T. J. McCabe. 1976. A Complexity Measure. *IEEE Transactions on Software Engineering* SE-2, 4 (1976), 308–320.

[66] Carma L. McClure. 1978. A Model for Program Complexity Analysis. In *Proceedings of the 3rd International Conference on Software Engineering* (Atlanta, Georgia, USA). 149–157.

[67] R. Mo, Y. Cai, R. Kazman, L. Xiao, and Q. Feng. 2016. Decoupling Level: A New Metric for Architectural Maintenance Complexity. In *2016 IEEE/ACM 38th International Conference on Software Engineering*. 499–510.

[68] Ran Mo, Shaozhi Wei, Qiong Feng, and Zengyang Li. 2022. An Exploratory Study of Bug Prediction at the Method Level. *Inf. Softw. Technol.* 144, C (apr 2022).

[69] Audris Mocku and Lawrence G. Votta. 2000. Identifying reasons for software changes using historic databases. In *Proceedings 2000 International Conference on Software Maintenance*. 120–130.

[70] A. Monden, D. Nakae, T. Kamiya, S. Sato, and K. Matsumoto. 2002. Software quality analysis by code clones in industrial legacy software. In *Proceedings IEEE Symposium on Software Metrics*. 87–94.

[71] Raimund Moser, Witold Pedrycz, and Giancarlo Succi. 2008. Analysis of the Reliability of a Subset of Change Metrics for Defect Prediction. In *Proceedings of the Second ACM-IEEE International Symposium on Empirical Software Engineering and Measurement* (Kaiserslautern, Germany) *(ESEM '08)*. 309–311.

[72] N. Nagappan and T. Ball. 2005. Use of relative code churn measures to predict system defect density. In *Proceedings. 27th International Conference on Software Engineering*. 284–292.

[73] Mohammed Amine Naji, Sanaa El Filali, Meriem Bouhlal, EL Habib Benlahmar, Rachida Ait Abdelouhahid, and Olivier Debauche. 2021. Breast cancer prediction and diagnosis through a new approach based on majority voting ensemble classifier. *Procedia Computer Science* 191 (2021), 481–486.

[74] Tsz Hin Ng, Shing-Chi Cheung, Wing Kwong Chan, and Yuen-Tak Yu. 2006. Toward effective deployment of design patterns for software extension: A case study. In *Proceedings of the 2006 international workshop on Software quality*. 51–56.

[75] Steffen M. Olbrich, Daniela S. Cruzes, and Dag I.K. Sjøberg. 2010. Are all code smells harmful? A study of God Classes and Brain Classes in the evolution of three open source systems. In *2010 IEEE International Conference on Software Maintenance*. 1–10.

[76] Julius Omona. 2013. Sampling in qualitative research: Improving the quality of research outcomes in higher education. *Makerere Journal of Higher Education* 4, 2 (2013), 169–185.

[77] Anthony J Onwuegbuzie and Nancy L Leech. 2007. A call for qualitative power analyses. *Quality & quantity* 41, 1 (2007), 105–121.

[78] Fabio Palomba, Gabriele Bavota, Massimiliano Di Penta, Fausto Fasano, Rocco Oliveto, and Andrea De Lucia. 2018. On the diffuseness and the impact on maintainability of code smells: a large scale empirical investigation. In *Proceedings of the 40th International Conference on Software Engineering*. 482–482.

[79] Fabio Palomba, Andy Zaidman, Rocco Oliveto, and Andrea De Lucia. 2017. An Exploratory Study on the Relationship between Changes and Refactoring. In *Proceedings of the 25th International Conference on Program Comprehension* (Buenos Aires, Argentina). 176–185.

[80] Fabio Palomba, Marco Zanoni, Francesca Arcelli Fontana, Andrea De Lucia, and Rocco Oliveto. 2017. Toward a smell-aware bug prediction model. *IEEE Transactions on Software Engineering* 45, 2 (2017), 194–218.

[81] Luca Pascarella, Fabio Palomba, and Alberto Bacchelli. 2020. On the performance of method-level bug prediction: A negative result. *Journal of Systems and Software* 161 (2020).

[82] Daryl Posnett, Abram Hindle, and Premkumar Devanbu. 2011. A Simpler Model of Software Readability. In *Proceedings of the 8th Working Conference on Mining Software Repositories* (Waikiki, Honolulu, HI, USA). 73–82.

[83] Aniket Potdar and Emad Shihab. 2014. An exploratory study on self-admitted technical debt. In *2014 IEEE International Conference on Software Maintenance and Evolution*. 91–100.

[84] Danijel Radjenović, Marjan Heričko, Richard Torkar, and Aleš Živkovič. 2013. Software fault prediction metrics: A systematic literature review. *Information and Software Technology* 55, 8 (2013), 1397 – 1418.

[85] Md Saidur Rahman and Chanchal K. Roy. 2017. On the Relationships Between Stability and Bug-Proneness of Code Clones: An Empirical Study. In *2017 IEEE 17th International Working Conference on Source Code Analysis and Manipulation (SCAM)*. 131–140.

[86] Paul Ralph and Ewan Tempero. 2018. Construct Validity in Software Engineering Research and Software Metrics. In *Proceedings of the 22nd International Conference on Evaluation and Assessment in Software Engineering 2018* (Christchurch, New Zealand). 13–23.

[87] Bruce Ratner. [n. d.]. The correlation coefficient: Its values range between+ 1/- 1, or do they? *Journal of targeting, measurement and analysis for marketing* 17, 2 ([n. d.]), 139–142.

[88] Baishakhi Ray, Vincent Hellendoorn, Saheel Godhane, Zhaopeng Tu, Alberto Bacchelli, and Premkumar Devanbu. 2016. On the "Naturalness" of Buggy Code. In *Proceedings of the 38th International Conference on Software Engineering* (Austin, Texas) *(ICSE '16)*. 428–439.

[89] Nornadiah Mohd Razali, Yap Bee Wah, et al. 2011. Power comparisons of shapiro-wilk, kolmogorov-smirnov, lilliefors and anderson-darling tests. *Journal of statistical modeling and analytics* 2, 1 (2011), 21–33.

[90] D. Romano and M. Pinzger. 2011. Using source code metrics to predict change-prone Java interfaces. In *2011 27th IEEE International Conference on Software Maintenance*. 303–312.

[91] Giovanni Rosa, Luca Pascarella, Simone Scalabrino, Rosalia Tufano, Gabriele Bavota, Michele Lanza, and Rocco Oliveto. 2021. Evaluating SZZ Implementations Through a Developer-Informed Oracle. In *Proceedings of the 43rd International Conference on Software Engineering* (Madrid, Spain). 436–447.

[92] S. Scalabrino, G. Bavota, C. Vendome, M. Linares-Vásquez, D. Poshyvanyk, and R. Oliveto. 2017. Automatically assessing code understandability: How far are we?. In *32nd IEEE/ACM International Conference on Automated Software Engineering*. 417–427.

[93] Simone Scalabrino, Mario Linares-Vásquez, Rocco Oliveto, and Denys Poshyvanyk. 2018. A comprehensive model for code readability. *Journal of Software: Evolution and Process* 30, 6 (2018), e1958.

[94] S. Scalabrino, M. Linares-Vásquez, D. Poshyvanyk, and R. Oliveto. 2016. Improving code readability models with textual features. In *IEEE 24th International Conference on Program Comprehension*. 1–10.

[95] Ingo Scholtes, Pavlin Mavrodiev, and Frank Schweitzer. 2016. From Aristotle to Ringelmann: a large-scale analysis of team productivity and coordination in Open Source Software projects. *Empirical software engineering : an international journal* 21, 2 (2016), 642–683.

[96] Francisco Servant and James A. Jones. 2017. Fuzzy Fine-Grained Code-History Analysis. In *Proceedings of the International Conference on Software Engineering (ICSE)*. 746–757.

[97] Kanwarpreet Sethi, Yuanfang Cai, Sunny Wong, Alessandro Garcia, and Claudio Sant'Anna. 2009. From retrospect to prospect: Assessing modularity and stability from software architecture. In *2009 Joint Working IEEE/IFIP Conference on Software Architecture & European Conference on Software Architecture*. 269–272.

[98] Emad Shihab, Ahmed E. Hassan, Bram Adams, and Zhen Ming Jiang. 2012. An Industrial Study on the Risk of Software Changes. In *Proceedings of the ACM SIGSOFT 20th International Symposium on the Foundations of Software Engineering* (Cary, North Carolina).

[99] Emad Shihab, Audris Mockus, Yasutaka Kamei, Bram Adams, and Ahmed E Hassan. 2011. High-impact defects: a study of breakage and surprise defects. In *Proceedings of the 19th ACM SIGSOFT symposium and the 13th European conference on Foundations of software engineering*. 300–310.

[100] Boris Shiklo and Andy Lipnitski. [n. d.]. Software Maintenance Costs in Brief . https://www.scnsoft.com/software-development/maintenance-and-support/costs. [Online; last accessed 18-June-2024].

[101] Y. Shin, A. Meneely, L. Williams, and J. A. Osborne. 2011. Evaluating Complexity, Code Churn, and Developer Activity Metrics as Indicators of Software Vulnerabilities. *IEEE Transactions on Software Engineering* 37, 6 (2011), 772–787.

[102] Thomas Shippey, Tracy Hall, Steve Counsell, and David Bowes. 2016. So You Need More Method Level Datasets for Your Software Defect Prediction? Voilà! *(ESEM '16)*.

[103] Danilo Silva, Nikolaos Tsantalis, and Marco Tulio Valente. 2016. Why we refactor? confessions of github contributors. In *Proceedings of the 2016 24th acm sigsoft international symposium on foundations of software engineering*. 858–870.

[104] Dag IK Sjøberg, Bente Anda, and Audris Mockus. 2012. Questioning software maintenance metrics: a comparative case study. In *Proceedings of the ACM-IEEE international symposium on Empirical software engineering and measurement*. 107–110.

[105] D. I. K. Sjøberg, A. Yamashita, B. C. D. Anda, A. Mockus, and T. Dybå. 2013. Quantifying the Effect of Code Smells on Maintenance Effort. *IEEE Transactions on Software Engineering* 39, 8 (2013), 1144–1156.

[106] D. Spadini, F. Palomba, A. Zaidman, M. Bruntink, and A. Bacchelli. 2018. On the Relation of Test Smells to Software Code Quality. In *2018 IEEE International Conference on Software Maintenance and Evolution*. 1–12.

[107] D. Ståhl, A. Martini, and T. Mårtensson. 2019. Big Bangs and Small Pops: On Critical Cyclomatic Complexity and Developer Integration Behavior. In *2019 IEEE/ACM 41st International Conference on Software Engineering: (ICSE-SEIP)*. 81–90.

[108] Daniela Steidl and Florian Deissenboeck. 2015. How do java methods grow?. In *2015 IEEE 15th International Working Conference on Source Code Analysis and Manipulation (SCAM)*. 151–160.

[109] Daniela Steidl, Benjamin Hummel, and Elmar Juergens. 2013. Quality analysis of source code comments. In *2013 21st international conference on program comprehension (icpc)*. 83–92.

[110] Daniela Steidl, Benjamin Hummel, and Elmar Juergens. 2014. Incremental Origin Analysis of Source Code Files. In *Proceedings Working Conference on Mining Software Repositories (MSR)*. 42—-51.

[111] Klaas-Jan Stol, Paul Ralph, and Brian Fitzgerald. 2016. Grounded theory in software engineering research: a critical review and guidelines. In *Proceedings of the 38th International conference on software engineering*. 120–131.

[112] Anselm Strauss and Juliet Corbin. 1998. Basics of qualitative research techniques. (1998).

[113] Zhongbin Sun, Junqi Li, Heli Sun, and Liang He. 2021. CFPS: Collaborative filtering based source projects selection for cross-project defect prediction. *Applied Soft Computing* 99 (2021), 106940.

[114] Chakkrit Tantithamthavorn and Ahmed E. Hassan. 2018. An Experience Report on Defect Modelling in Practice: Pitfalls and Challenges. In *Proceedings of the 40th International Conference on Software Engineering: Software Engineering in Practice* (Gothenburg, Sweden). 286–295.

[115] Qingye Wang, Xin Xia, David Lo, and Shanping Li. 2019. Why is my code change abandoned? *Information and Software Technology* 110 (2019), 108 – 120.

[116] Song Wang, Junjie Wang, Jaechang Nam, and Nachiappan Nagappan. 2021. Continuous Software Bug Prediction. Article 14, 12 pages.

[117] Tzu-Tsung Wong. 2015. Performance evaluation of classification algorithms by k-fold and leave-one-out cross validation. *Pattern recognition* 48, 9 (2015), 2839–2846.

[118] Pavlína Wurzel Gonçalves, Gül Calikli, Alexander Serebrenik, and Alberto Bacchelli. 2023. Competencies for code review. *Proceedings of the ACM on Human-Computer Interaction* 7, CSCW1 (2023), 1–33.

[119] Meng Yan, Xin Xia, Emad Shihab, David Lo, Jianwei Yin, and Xiaohu Yang. 2018. Automating change-level self-admitted technical debt determination. *IEEE Transactions on Software Engineering* 45, 12 (2018), 1211–1229.

[120] Xiaofang Zhang, Yida Zhou, and Can Zhu. 2017. An empirical study of the impact of bad designs on defect proneness. In *2017 International conference on software analysis, testing and evolution (SATE)*. 1–9.

[121] Yuming Zhou, Baowen Xu, and Hareton Leung. 2010. On the ability of complexity metrics to predict fault-prone classes in object-oriented systems. *Journal of Systems and Software* 83, 4 (2010), 660 – 674.

[122] Thomas Zimmermann, Rahul Premraj, and Andreas Zeller. 2007. Predicting Defects for Eclipse. In *Proceedings of the Third International Workshop on Predictor Models in Software Engineering*. 7 pages.